\documentclass[aps,prx,reprint,amsfonts,amsmath,amssymb,groupedaddress,superscriptaddress,onecolumn]{revtex4-2}
\usepackage{graphicx}
\usepackage{makecell}
\usepackage{bm}
\usepackage[colorlinks,urlcolor=blue,anchorcolor=blue,linkcolor=blue,citecolor=blue,breaklinks=true]{hyperref}

\begin{document}
\title{Potential-tuned magnetic switches and half-metallicity transition in zigzag graphene nanoribbons}
\date{\today}

\begin{abstract}
Realizing controllable room-temperature ferromagnetism in carbon-based materials is one of recent prospects. The magnetism in graphene nanostructures reported previously is mostly formed near the vacancies, zigzag edges, or impurities by breaking the local sublattice imbalance, though a bulk chiral spin-density-wave ground state is also reported at van Hove filling due to its perfectly nested Fermi surface. Here, combining of the first-principles and tight-binding model simulations, we predict a robust ferromagnetic domain lies between the inter-chain carbon atoms inside the zigzag graphene nanoribbons by applying a potential drop. We show that the effective zigzag edges provide the strong correlation background through narrowing the band width, while the internal Van Hove filling provides the strong ferromagnetic background inherited from the bulk. The induced ferromagnetism exhibit interesting switching effect when the nominal Van Hove filling crosses the intra- and inter-chain region by tuning the potential drops. We further observe a robust half-metallicity transition from one spin channel to another within the same magnetic phase. These novel properties provide promising ways to manipulate the spin degree of freedom in graphene nanostructures.
\end{abstract}

\author{Wei-Jian Li}
\affiliation{National Laboratory of Solid State Microstructure, Department of
Physics, Nanjing University, Nanjing 210093, China}
\author{Shi-Chang Xiao}
\affiliation{National Laboratory of Solid State Microstructure, Department of
Physics, Nanjing University, Nanjing 210093, China}
\author{Da-Fei Sun}
\affiliation{National Laboratory of Solid State Microstructure, Department of
Physics, Nanjing University, Nanjing 210093, China}
\author{Chang-De Gong}
\affiliation{Center for Statistical and Theoretical Condensed Matter
Physics, Zhejiang Normal University, Jinhua 321004, China}
\affiliation{National Laboratory of Solid State Microstructure, Department of
Physics, Nanjing University, Nanjing 210093, China}
\affiliation{Collaborative Innovation Center of Advanced Microstructures, Nanjing University, Nanjing 210093, China}
\author{Shun-Li Yu}
\email{slyu@nju.edu.cn}
\affiliation{National Laboratory of Solid State Microstructure, Department of Physics, Nanjing University, Nanjing 210093, China}
\affiliation{Collaborative Innovation Center of Advanced Microstructures, Nanjing University, Nanjing 210093, China}
\author{Yuan Zhou}
\email{zhouyuan@nju.edu.cn}
\affiliation{National Laboratory of Solid State Microstructure, Department of Physics, Nanjing University, Nanjing 210093, China}
\affiliation{Collaborative Innovation Center of Advanced Microstructures, Nanjing University, Nanjing 210093, China}
\maketitle

The advances in atomically precise fabrication synthesized by the bottom-up routes through molecular precursors in graphene nanostructures with variety of edge geometries, sizes and dopants~\cite{CaiJ-Nature2010,Ruffieux-Nature2016,Bronner-ACSN2018,ChenZP-AM2020}, open the prospective platform to engineer their novel physical properties. The graphene nanoribbons, including the armchair nanoribbons (AGNRs), zigzag nanoribbons (ZNGRs), host widely tunable band structures. The band gap in the AGNRs is readily engineered by width~\cite{Diez-ACSN2017}, external electric field~\cite{Pizzochero-NL2021}, as well as their heterojunctions~\cite{Abbassi-ACSN2020,Cernevics-PRB2020}. According to the number of carbon atoms acrossing the ribbon~\cite{Son-PRL2006,Yazyev-ACR2013}, the AGNRs exhibit the intriguing semiconductor-semimetal transition and band topology~\cite{CaoT-PRL2017,Rizzo-Nature2018,Groning-Nature2018}. In contrast, the ZGNRs usually develop the spin-polarized edge states~\cite{Slota-Nature2018,LiJC-NC2019} and the character of half-metallicity~\cite{Son-Nature2006}. In particular, the long-distance and long-time nature of the spin relaxation in graphene~\cite{Tombros-Nature2007,Herrero-Science2016} enable the ZGNRs to be an ideal platform for spintronics~\cite{Pesin-NM2012}.

Essentially, the pristine graphene is a broadband semimetal with a bandwidth in the order of $16$ eV~\cite{Neto-RMP2009}, which makes the electron correlation to be weak and magnetism not to naturally appear. Thus, in order to realize the magnetism in graphene, the correlated effect should be enhanced. In this respect, narrowing the bandwidth to quench the kinetic energy of electrons is an effective route. For example, the twisted bilayer graphene has recently emerged as a highly tunable platform to realize low-energy narrow bands for the experimental investigation of many strongly correlated phases~\cite{Bistritzer-PNAS2011,Kim-PNAS2017}, such as Mott physics and superconductivity~\cite{Cao-Nature2018-1,Cao-Nature2018-2}. Similar strategy can also be achieved by creating atomic vacancies/defects~\cite{Cervenka-NP2009,Zhang-PRL2016}, absorbing adatoms~\cite{Herrero-Science2016,Zheng-PRL2020} and elemental substitutions~\cite{Tucek-AM2016,Blonski-JACS2017,Babar-PRB2019,Di-NJP2021} to induce the localized states and net magnetic moment in this respect. The strong low-temperature ferromagnetism, and even the near room-temperature ferromagnetism have been reported~\cite{Liu-NC2016,Fu-ACSN2019,Fu-PRB2020}. The zigzag-edge graphene nanostructures is another typical example\cite{Rossier-PRL2007,Hu-PRB2012}. Removing the free p$_{z}$ orbital from $\pi$-graphene system creates a partially flat band at the Fermi level in the ZNGRs~\cite{Fujita-JPSJ1996} and therefore enhances the correlated effects. Strong spin polarization on the boundaries with antiparallel orientation on the opposite edges is found in the graphene ribbons~\cite{Hu-JPCC2014,Ma-PRB2016}, leading to a giant edge state splitting~\cite{WangSY-NC2016}. The parallel spin polarization on the boundaries due to the quantum interference is further reported in the wider graphene nanoribbons with slight charge doping~\cite{Magda-Nature2014,Chen-NL2017}. The magnetism generated above is mainly localized, and can be attributed to the breaking the local sublattice imbalance according to the Lieb's theorem on bipartite lattice~\cite{Lieb-PRL1989}. Besides the local origin of magnetism, we can also take advantage of its special electronic structure at certain doping levels to realize magnetism in graphene. In particular, the $\frac{3}{4}$-filled graphene is predicted to be in the chiral spin density wave state~\cite{TLi-EPL2012,Wang-PRB2012} at zero temperature, and the uniaxial spin density wave state~\cite{Nandkishore-PRL2012} at low temperature due to the perfect Fermi surface nesting and the intensive density of states originating from van Hove singularities. Despite these quick improvements, the efficient and precise controlling of the ferromagnetism remain challenge in carbon-based materials, in particular inside the nanostructures.

Here, we report the robust potential-tuned ferromagnetic switches and half-metallicity transitions on the zigzag graphene nanoribbons in combination of the first-principles calculations and the model simulations. Applying the potential drops, the ferromagnetic domains steadily lies between the inter-chain carbon atoms when the nominal Van hove filling crosses them. We show that the zigzag edges and the inter-chain Van Hove fillings provide the large density of states, strong tendency of instability towards the ferromagnetism, and strong interaction, which are the essential ingredients for the ferromagnetically ordered state. The Van hove filling moves inside between the inter-chain and intra-chain alternatively by strengthening the potential drops, realizing the switches of ferromagnetic domains. We further predict a half-metallicity alternation from one to another spin channel within the same magnetic state. Our results offer a new pathway to manipulate the spin degree of freedom in future graphene nanostructures.

\begin{figure*}[tbp!]
\center
\includegraphics[width=\textwidth]{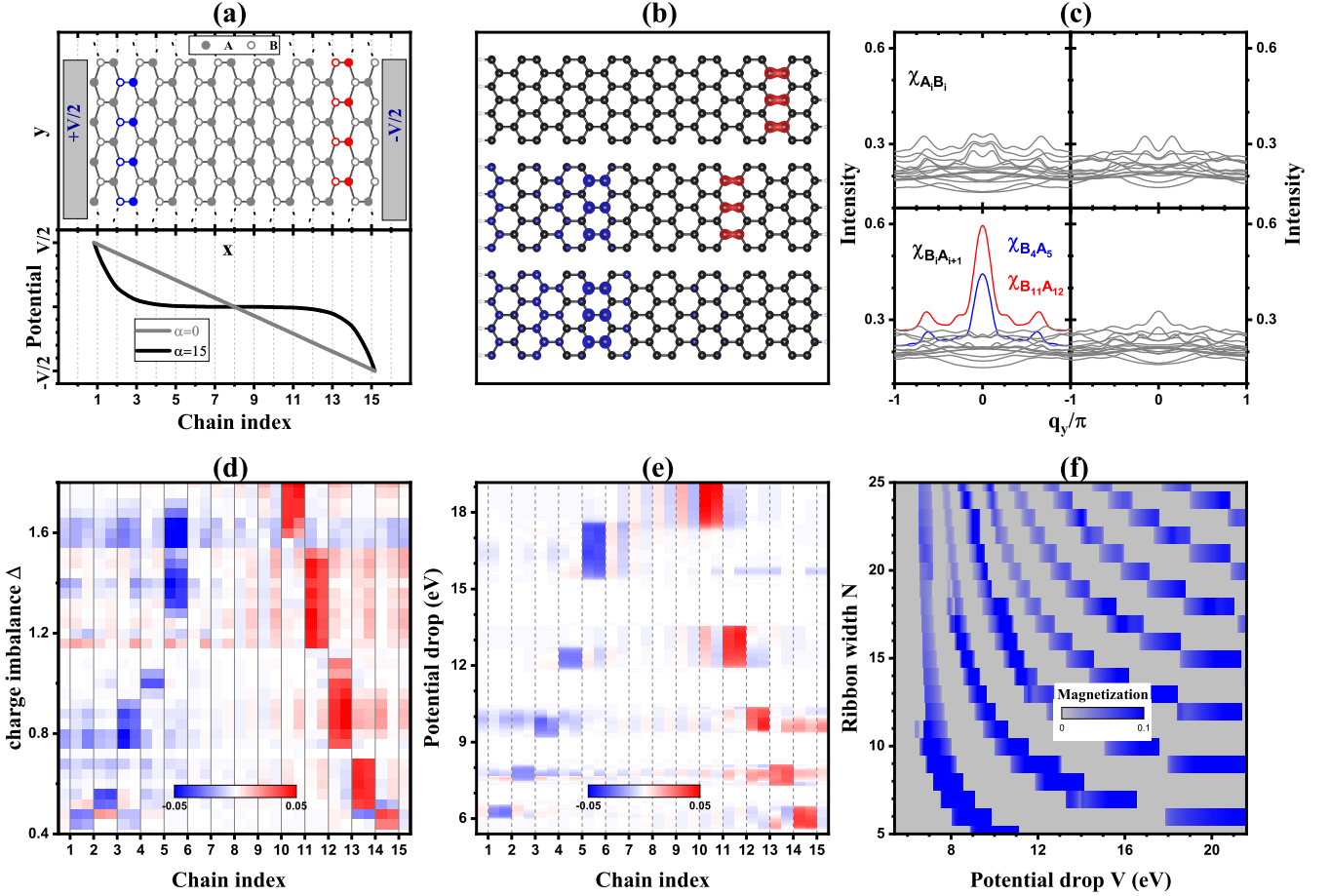}
\caption{(a) Schematic $15$-ZGNR in potential field. Upper panel: $15$-ZGNR with a fixed potential drop along the $x$-direction. The ferromagnetic domains in the $2^{\text{nd}}$ inter-chain state is highlighted (blue and red symbols). Low panel: the schematic potential field, $\alpha=0$ for homogeneous potential field, and $\alpha=15$ for the strongly attenuated potential field. (b) Distribution of magnetism in the $i^{\text{th}}$ inter-chain state simulated by the first-principles calculations, from top to bottom, $\Delta=0.6$, $1.4$, and$1.56$, respectively the isosurface value of the spin-polarized electron density is $0.003$ $e$/\AA$^{3}$, blue and red for spin-down and spin up, respectively. (c) Static spin susceptibility in respective channels in the non-interacting system, left for the $4^{\text{th}}$ inter-chain state with $V=12.5$ eV and right for the $4^{\text{th}}$ intra-chain state with $V=11.5$ eV. From top to bottom, data are the AB ($\chi_{A_{i}B_{i}}$), and BA ($\chi_{B_{i}A_{i+1}}$) channels, respectively. (d) Magnetic phase diagram in the potential-tuned $15$-ZGNR simulated by first-principles calculations, and (e) similar magnetic phase diagram simulated by the extend-$\pi$-orbital Hubbard model. (f) Width-potential ($N-V$) magnetic phase diagram simulated by the simple-$\pi$-orbital Hubbard model. The magnitude of magnetization shown here is the summation over the half-width ribbon (left) with $M_{t}=\sum_{i\in\text{left}}M_{i}$. The homogeneous potential drops ($\alpha=0)$ is assumed in all figures.}
\label{Mag}
\end{figure*}

The considered system is a $N$-coupled zigzag graphene nanoribbon (N-ZGNR) in the potential field. The applied potential is assumed to linearly ($\alpha=0$) or exponentially decay ($\alpha> 0$) from the boundaries into the bulk with the fixed potential drops $V$ between the two terminations (Fig.~\ref{Mag} (a)). We focus on the strong field $5<V<20$ (eV) situation, corresponding to about $1.6 \sim 6.6$ eV/nm for the $15$-ZGNR, which is comparable to the previous applied electric field on AGNRs~\cite{Pizzochero-NL2021}. To reveal the magnetic properties of the potential-tuned ZGNRs, we consider both the first-principles and the tight-binding model simulations. The first-principles simulation is performed in the framework of the density function theory using the virtual crystal approximations. In practice, the initial charge imbalance $\Delta$ defined by the charge difference between the two terminations and a linear dependence in bulk are assumed, each $0.1$ of the charge imbalance roughly corresponds to the potential drop $6\sim 8$ eV (See Methods and supplementary~\ref{S1B} for details). The adopted tight-binding models are the extend-$\pi$-orbital Hubbard model with the typical tight-binding parameters $t=2.7$ eV, $t^{\prime}=0.2$ eV, and $t^{\prime\prime}=0.18$ eV~\cite{Kundu-MPLB2011}, and the simple-$\pi$-orbital Hubbard model with only the nearest-neighbor hopping $t=2.7$ eV (See Models for details). The on-site Coulomb repulsion is set as $U=2t$ unless specified, and the temperature in the model simulations is fixed at near room-temperature with $T=0.027$ eV. We have checked the magnetic properties, as well as the metallicity in the low fields. The dominating magnetization robustly localized along the boundaries with the antiferromagnetically correlated edges and an insulating to half-metallic phase transition under low field are observed (Supplementary~\ref{S1A}). These features well reproduce the previous results found experimentally~\cite{Magda-Nature2014} or predicted theoretically~\cite{Son-Nature2006,Son-PRL2006}, manifesting the validity of present methods and models.\bigskip

\noindent\textbf{Results}\\
\textbf{Ferromagnetic domains and magnetic switches.} The edge magnetization is unstable due to the heavy carrier doping on the edges in the stronger potential field. Unexpectedly, the first-principles simulations reveal that the robust inter-chain ferromagnetic domains develop in the bulk of the ribbon in certain fields region as shown in the phase diagram of $15$-ZGNR in Fig.~\ref{Mag}(d), i.e., the ferromagnetism is mainly concentrated on the neighboring inter-chain $B_{i}$ and $A_{i+1}$ sublattices as depicted in Fig.~\ref{Mag} (b), in analogy to the local configuration of the stripy phase found in the Heisenberg-Kitaev spin model~\cite{Chaloupka-PRL2013}. The present local ferromagnetism exhibit the typical $\pi$-orbital magnetism, which also supports our $\pi$-orbital model simulations. We define the $i^{\text{th}}$ inter-chain state as the ferromagnetic domain dominating the $B_{i}$ and $A_{i+1}$ sites (and their counterpart $A_{N+1-i}$ and $B_{N-i}$ in electron-doped side), and subsequently the $i^{\text{th}}$ intra-chain state as a paramagnetic state with the slightly low potential drop as schematically shown in Fig.~\ref{Mag} (a). The present inter-chain ferromagnetism is in sharp contrast to the previous magnetism reported in low electric fields, which manifests itself as the strong edge spin polarization but with weakly antiferromagnetic correlations between the inter- and intra-chain ($A_{i}$ and $B_{i}$) sublattices~\cite{Jiang-JCP2007,Magda-Nature2014,Ma-PRB2016}. The difference between the ferromagnetic domain and antiferromagnetic edge magnetization resembles the situation in two-dimensional Hubbard model on the square lattice, where the system favors antiferromagnetic state near half-filling, but ferromagnetic states in heavy doping~\cite{Lin-PRB1987}. The magnetic phase diagrams simulated by the extend/simple-$\pi$-orbital Hubbard model under the mean-field approximation well support the emergence of the ferromagnetic domains as shown in Fig.~\ref{Mag} (e) and (f). The qualitative agreement between the first-principles and model simulations indicates the intrinsic nature of magnetic properties in the potential-tuned ZGNRs.

The $i^{\text{th}}$ inter-chain ferromagnetic domains in the electron- and hole-doped sides are antiferromagnetically correlated. It is more evident in the simple-$\pi$-orbital model simulations with only the nearest-neighbor hopping included. In this situation, the particle-hole symmetry is strictly preserved, and the antiferromagnetic correlation between the ferromagnetic domains in the electron- and hole-doped sides obeys the Lieb's theorem in bipartite lattice~\cite{Lieb-PRL1989}. In contrast, the particle-hole symmetry breaks in the realist graphene nanoribbons due to the presence of the long-range hopping process. As a consequence, the ferromagnetic domains in both sides are no more coincident as revealed by both the first-principle and extend-$\pi$-orbital Hubbard simulations, in particular, under the stronger field. We note that the ferromagnetic domains in the electron-doped side are more robust than that in the hole-doped side, in agreement with the larger density of states at the Van Hove filling in the electron-doped graphene~\cite{Neto-RMP2009}.

The inter-chain ferromagnetism is also supported by the non-interacting static spin susceptibility, also known as the Linhard function, $\chi_{\eta_{i}\eta^{\prime}_{j}}(q_{y})$ with $\eta=A/B$ the sublattice index and $i/j$ the chain index. We show the two representative results in Fig.~\ref{Mag} (c). In the $4^{\text{th}}$ inter-chain ($V=12.5$ eV), the prominent peaks with positive values are found in the channels of $\chi_{A_{5}A_{5}}$, $\chi_{B_{4}B_{4}}$, and $\chi_{B_{4}A_{5}}$ at $q_{y}=0$ (also their counterparts in electron-doped side), indicating the strong tendency towards the ferromagnetic instability between $4^{\text{th}}$ inter-chain sites. Instead, the weak or negligible peaks are observed in other channels, manifesting the weak magnetic correlations betweens those nearby chains. The potential-tuned ZGNR is, therefore, expected to develop the local inter-chain ferromagnetic domains when
the finite Coulomb interaction $U$ is included. In contrast, no significant peak is found in the $4^{\text{th}}$ intra-chain states at $V=11.5$ eV, corresponding to a paramagnetic state (also see Supplementary~\ref{S3} for details). For completeness, we also present the results in the absence of the potential drop. The significant positive peak at $q_{y}=0$ is found in $\chi_{A_{1}A_{1}}$ (and $\chi_{B_{15}B_{15}}$ channels (Supplementary~\ref{S3}), while the negligible peaks in the other channels. These features well agree with the strong spin polarization along the edges as revealed in the previous studies~\cite{Jiang-JCP2007,Ma-PRB2016,Chen-NL2017}. Therefore, the analysis on static spin susceptibilities supports the observed ferromagnetic domains and the magnetic switches in the potential-tuned ZNGRs, which is significantly differs from the edge magnetization reported before~\cite{Magda-Nature2014}.

The ferromagnetic domains are well engineered by the applied potential drops. They gradually move into the bulk but steadily locate between the inter-chain sites, undergoing an intermediate paramagnetic state (intra-chain state), when the potential drop is enhanced. As a consequence, the potential-tuned switches of the ferromagnetism is realized. The ferromagnetic domains, as well as the switches, are also readily tuned by the width of ribbons (Fig.~\ref{Mag} (f)). The phase alternations between the magnetic and paramagnetic phases are more frequent in the wider ribbons. The observed magnetic switches are robust against the detailed potential decay ratio ($\alpha$), the ribbon width ($N$), and the on-site Coulomb interaction $U$ (Supplementary~\ref{S1B}), again manifesting the intrinsic nature of the graphene ribbon under the strong field. The robustness of ferromagnetic domains and easy manipulations provide new opportunity to control the bulk magnetism in ZGNRs.

\begin{figure*}[tbp!]
\vspace{-0in} \hspace{-0.0in} \center
\includegraphics[width=\textwidth]{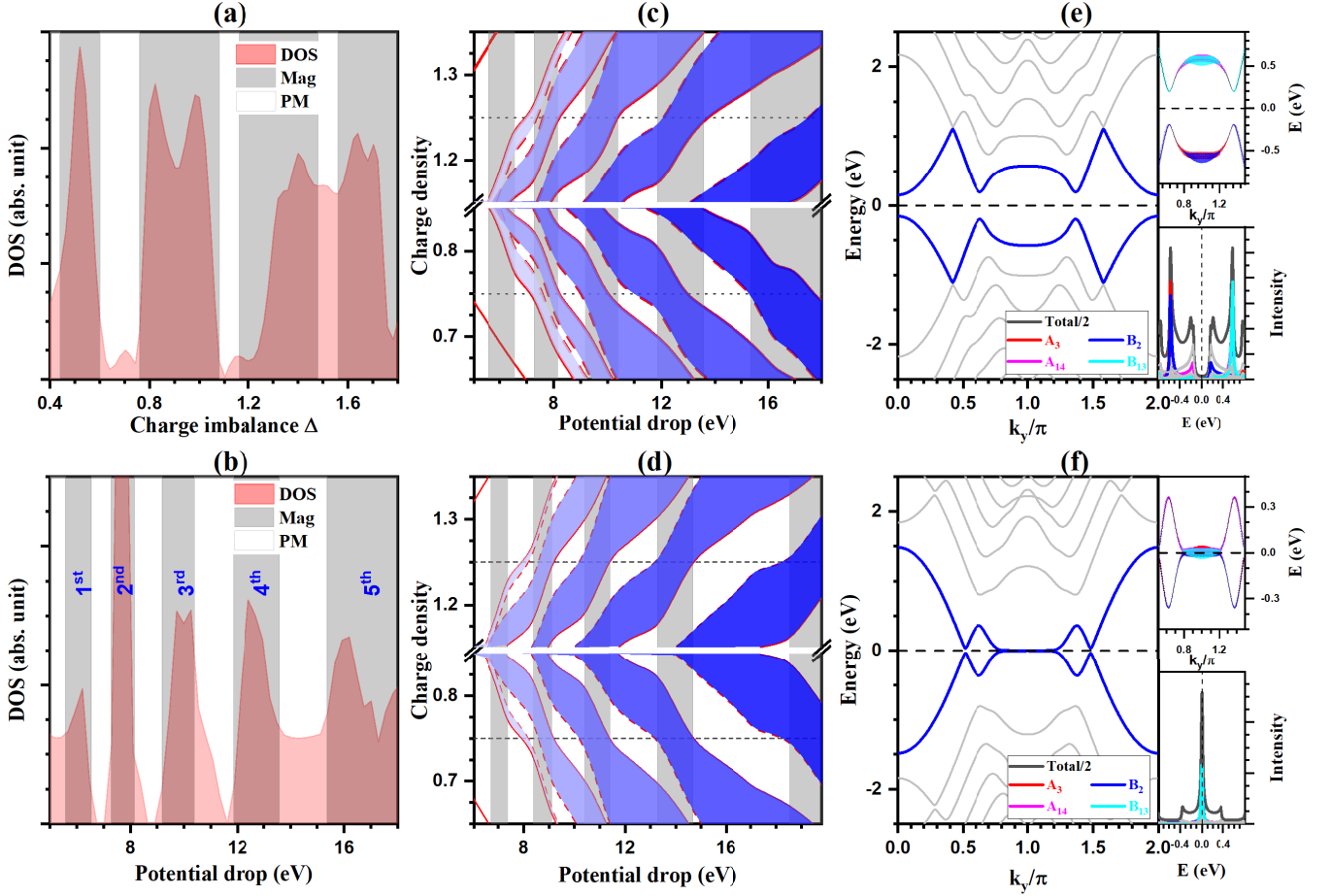}
\caption{(a) and (b), potential-tuned density of states, simulated by the first-principles calculations and the extend $\pi$-orbital Hubbard model. Here, the density of states is summed over a near Fermi energy interval $[-40,40]$ meV, integrated near the $k_{y}=\pi$ region ($k_{y}\in[\pi/2,3\pi/2]$). The $i^{\text{th}}$ inter-chain state (gray region) is highlighted by the colored notation, which corresponds to the colored region shown in (c) and (d). (c) and (d) are the charge distribution of respective sites simulated by the extend $\pi$-orbital and simplified $\pi$-orbital Hubbard model, respectively. From left to right are $A_{1}$ to $A_{6}$ (solid lines) and $B_{1}$ to $B_{5}$ (dash lines) in the lower panels, and the corresponding counterparts in upper panels. The color-shaded region highlights the inter-chain carbon sites ($1^{\text{st}}$ to $5^{\text{th}}$ from left to right). The dotted lines show the Van Hove filling of the bulk graphene. The homogeneous potential drops $(\alpha=0)$ are adoped in (a)-(d). (e)-(f) are the two typical low energy band structure for the $2^{\text{nd}}$ intra-chain state ($V=16.2$ eV)  and the $2^{\text{nd}}$ inter-chain state ($V=21.3$ eV) in the normal state of $15$-ZGNR with strongly decayed potential ($\alpha=10$), respectively. Two inserts in each panel are the weight of the lowest valence and highest conduction bands highlighted in left panels, and the density of states contributed from the respective sites, as well as the total density of state. Here, the simplified $\pi$-orbital model is adopted for simplicity.}
\label{Charge}
\end{figure*}

In principle, the large density of states, strong enough interactions, and tendency towards the specific instability are three key ingredients for a specifically ordered state. To unveil the origin of the inter-chain ferromagnetic domains and the magnetic switches in the potential-tuned ZGNRs, we turn to analyse those ingredients. The consistence between the results of non-interacting spin susceptibility and the first-principles simulations, as well as the self-consistent mean field solutions, suggests that we can understand the physics behind the discovered magnetic features by studying the low energy band structure in normal state. We show the density of states integrated near $k_{y}=\pi$ region ($k_{y}\in[\pi/2, 3\pi/2]$) within a near-Fermi energy interval of $[-40, 40]$ meV in normal state (Fig.~\ref{Charge} (a) and (b)). Apparently, the inter-chain ferromagnetic states, and the intra-chain paramagnetic states, emerge around the peaks and valleys of the density of states, respectively. In general, the applied potential drops directly modify the spatial charge distributions along the finite direction of ZGNRs. The evolution of the charge density at each site is further shown in Fig.~\ref{Charge} (c) and (d), which reveals that the magnetic features are closely related to the Van Hove filling of the bulk graphene, in particular in the simple-$\pi$-orbital Hubbard model simulations. The inter-chain ferromagnetic domain emerges when the Van Hove filling crosses the inter-chain carbon sites $B_{i}$ and $A_{i+1}$ (or their counterparts), whereas the situation of the Van Hove filling crossing the intra-chain carbon atoms yields the paramagnetic state. It is well-known that the ground state of the bulk graphene at the Van Hove filling favors the chiral spin-density-wave state at zero temperature~\cite{TLi-EPL2012,Wang-PRB2012} and the uniaxial spin-density-wave state at finite temperatures~\cite{Nandkishore-PRL2012}. In particular, the latter spin-density wave state hosts the ferromagnetic correlations between the inter-chain sites. Therefore, the inter-chain Van Hove filling provides both the large density of states and the strong inter-chain ferromagnetic fluctuations, in agreement with our previous analysis of the static spin susceptibility.

We show two typical low-energy band structures of the $2^\text{nd}$-intra- and $3^\text{rd}$ inter-chain states in Fig.~\ref{Charge} (e) and (f). Two narrow bands (valence and conduction band) near the Fermi level in the vicinity of $k_{y}=\pi$ are observed, substantially enhancing the strongly correlated effects in ZGNR. In particular, a nearly flat band is found in the $2^{\text{nd}}$ inter-chain state for the strong potential decay ratio ($\alpha=10$) at $V=21.3$ eV (Fig.~\ref{Charge} (f)), which well resembles to the partial flat band found in the absence of the field\cite{Neto-RMP2009}. The distinction is that the flat bands are dominated by the inter-chain atoms $B_{2}$ and $A_{3}$ (and their particle-hole symmetric counterparts) in the presence of potential drop (insert in Fig.~\ref{Charge} (f)), whereas they are only contributed by the edge atoms in the absence of potential drop. In contrast, the near Fermi level band also mainly comes from the inter-chain atoms $A_{3}$ and $B_{2}$ but with a substantial bandgap in the intra-chain state (Fig.~\ref{Charge} (e)) (More results are shown in Supplementary~\ref{S1B}). If we tentatively consider the narrow band near $k_{y}=\pi$ region as an edge-like state \emph{but} associated with the Van Hove fillings, the present band structures indicate a virtual edge moves into the bulk. The difference between the inter- and intra-chain state is that the virtual edge is a zigzag-edge in the former, while a dangling-edge in the latter. As well known that the dangling-edge graphene ribbon opens band gap near $k_{y}=\pi$ region, in sharp contrast with the gapless band in the zigzag-edge ribbons (Supplementary~\ref{S1B}). As a consequence, it naturally yields the 'U'- or 'V'-shaped density of states in the intra-chain, while the sharp peak at Fermi level in the inter-chain state (Fig.~\ref{Charge} (e) and (f)). In this sense, the zigzag edges of the ZGNRs provide both the large density of states and the strong interactions. Therefore, the inter-chain ferromagnetism originates both the zigzag edges and the bulk Van Hove filling, which provide three above mentioned key ingredients for the ferromagnetically ordered state.\bigskip

\begin{figure*}[tbp]
\vspace{-0in} \hspace{-0.0in} \center
\includegraphics[width=\textwidth]{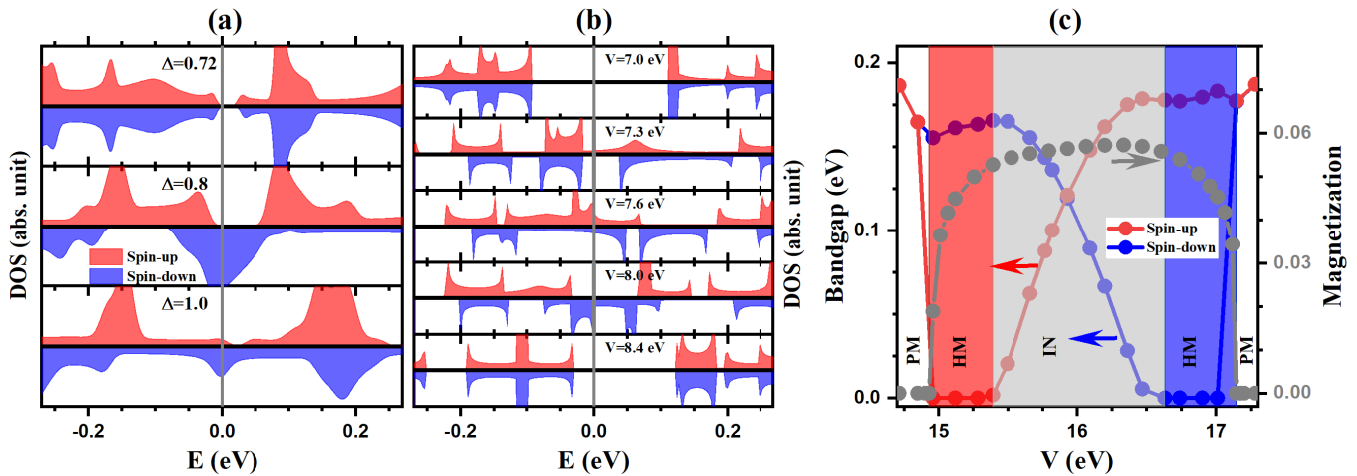}
\caption{(a) and (d) Spin resolved density of states simulated by the first-principles calcualtion and the extend $\pi$-orbital Hubbard model. Here, the density of states is an integration over $k_{y}\in[\pi/2, 3\pi/2]$ region. In (a), the upper panel is for the $2^{\text{nd}}$ intra-chain state ($\Delta=0.72$), and the lower two panels are for the $3^{\text{rd}}$ inter-chain state ($\Delta=0.8$ and $\Delta=1.0$). In (b), the top panel, and bottom panel in (b) is in the $2^{\text{nd}}$ intra-chain, and $3^{\text{rd}}$ intra-chain states, while the three middle panels are in the $2^{\text{nd}}$ inter-chain state, respectively. (c) The band gap (excluding the Dirac points) for respective spin channel simulated by the simple-$\pi$-orbital model in the $30$-ZGNR. The color region is the $2^{\text{nd}}$ inter-chain state. The gray circles is the magnetization summed over the half-width ribbon. Here, strong decay ratio $\alpha=15$ is assumed.}
\label{Halfmetal}
\end{figure*}

\noindent\textbf{Half-metallicity and half-metallic transition.} Since the emergence of the ferromagnetic domains breaks the local $SU(2)$ symmetry, it is natural to expect the emergence of the inequivalent transport between the two spin channels. Fig.~\ref{Halfmetal} (a) and (b) show the typical spin-resolved density of states integrated around $k_{y}=\pi$ in a small energy window near the Fermi energy. We find that the half-metal feature emerges at some potential drops in the magnetic state, i.e., the band gap closes for one spin-channel but opens for another spin-channel. We emphasize that the half-metallicity mentioned here is more precisely half-metal-like, since the dispersion near the Dirac points stemming from the bulk of graphene can be gapless (Supplementary~\ref{S2}).

Interestingly, we find a robust half-metallicity transition in the same magnetic region. Fig.~\ref{Halfmetal} (b) show a typical evolution of the spin-resolved density of states from the $2^{\text{nd}}$ intra-chain state to the $3^{\text{rd}}$ intra-chain state, simulated by the extend-$\pi$-orbital Hubbard model with the homogeneous potential drops (the detailed low-energy band structure is shown in Supplementary~\ref{S2}). A significant gap opens in both the spin-up and spin-down channels (degenerate) in the $2^{\text{nd}}$ intra-chain state (paramagnetic) at $V=7.0$ eV, yielding an insulating behavior. A half-metal feature with gap closing in the spin-up channel gradually develops when the ZGNR system enters the $2^{\text{nd}}$ inter-chain ferromagnetic state at $V=7.3$ eV. Further increasing the potential drop, both gaps of the spin-up and spin-down channels close, producing the metallic behavior. The half-metal feature emerges again but with the band gap opening in the reversed spin channel (spin-up) at $V=8.0$ eV. The ribbon system exhibits the insulating behavior after entering the $3^{\text{rd}}$ intra-chain paramagnetic state at $V=8.4$ eV. Therefore, we find a half-metallicity transition in the same inter-chain ferromagnetic state. The present half-metallicity, especially the metallicity transition between the reversal spin channels, is significantly distinct from the previously reported half-metallicity in the low electric field, where the metallicity is fixed in the same spin channel\cite{Son-Nature2006}. In fact, in consideration of the virtual edges between the inter-chain sites and the half-metal-to-insulator transition between the inter-chain ferromagnetic and intra-chain paramagnetic state, the insulator to half-metal transition predicted in the weak field~\cite{Son-Nature2006} is half of the process observed here. Such half-metallicity transition within the same magnetic phase is robust against the width of ribbon, and even the detailed potential drops. We further show similar process on the wide ribbon ($N=30$) with strong decay ratio ($\alpha=15$) simulated by the simplified-$\pi$-orbital Hubbard model. Due to the presence of particle-hole symmetry, the gap at Dirac points remain closed. The half-metal-like transition is also evident as shown in Fig.~\ref{Halfmetal} (c), where the bandgap is defined as the energy difference between the lowest valence and highest conduction bands but not including the small region near the Dirac points. We therefore predict a new path to engineer the electronic spin degree of freedom on graphene at the nanometre scale.\bigskip

\noindent\textbf{Discussion.}\\
Introducing the charge imbalance by the potential, we predict the potential-tuned inter-chain magnetic state in zigzag graphene nanoribbons. The bulk Van Hove filling and zigzag edges in the inter-chain state provide all requirements for the ferromagnetic state, i.e., the large density of states, strong tendency towards ferromagnetic ground state, and strong enough interactions. By tuning the potential drop, the ferromagnet-to-paramagnetic switches is realized. We further predict a robust half-metallicity transition between the reversed spin channels in the same inter-chain magnetic state. The controllability, robustness of the reported ferromagnetism and the half-metallicity transition provide new path to manipulate the spin degree of freedom in graphene nanostructures, and are prospective to the future spintronics.

To realize the suggested potential field, a simple consideration is the electric field as suggested before in low field case\cite{Son-Nature2006}. However, the strong screening effect of the electric field in metallic state naturally yields large decay ratio $\alpha$, impeding the charge modulations in bulk of ribbons. Such screening may substantially suppress the magnetic switches unless narrow enough ribbon is considered. On the other hand, the previously on-surface synthesis of graphene nanoribbons is usually based on the metal surface~\cite{CaiJ-Nature2010}, which further
screens the electronic properties of the designed graphene nanoribbons. Recently, the synthesis of graphene nanoribbons on the insulating and semiconducting substrates, such as the semiconducting metal oxide TiO$_{2}$~\cite{Kolmer-Science2020} and insulating h-BN~\cite{Wang-NM2021,Ilyasov-JAP2015} or SiC~\cite{Aprojanz-NC2018}, has been developed. They provide the opportunity to observe the magnetic property predicted here by applying the field on the insulating/semiconducting substrate to realize the desired potential.

Doping in the ZGNRs further breaks the particle-hole symmetry in ZGNRs. As discussed above, the local magnetic momentum emerges when the Van Hove filling crosses the inter-chain atoms ($A_{i+1}$ and $B_{i}$ or their counterparts). The emergence of ferromagnetic domains for the hole- and electron-side will be more asynchronous, offering more freedom to manipulate the bulk magnetism in potential-tuned ZGNR.
\bigskip

\noindent\textbf{Methods}\\
\textbf{First-principles simulations.} To explore the magnetic properties of the potential-tuned ZGNR, the first-principles calculations in the framework of the density function theory implemented in the Quantum Espresso package is performed~\cite{Giannozzi_JPCM2009}. We use the Perdew-Burke-Ernzerhof type generalized gradient approximation~\cite{Blochl-PRB1994} by using scalar relativistic and norm-conserving pseudopotentials of carbon and hydrogen with plane-wave cutoff of $90$ Ry~\cite{Setten-CPC2018}. $k$-point grids are taken as ($\Gamma$ centered) $14\times 1\times 1$ for the $15$-ZGNR. Vacuum layer with $15$ \AA \ thickness is used in $x$- and $z$-direction to ensure the decoupling between neighboring slabs and zigzag shaped edges on both sides passivated by hydrogen atoms. We consider the simplest homogeneous potential drop by using the virtual crystal approximations in our ab initio simulations. A relative `weight' of carbon atoms $w_{i}$ in $15$-ZGNR is set by a gradient with their values corresponding to the effective applied linear potential drop. We have checked the magnetic properties, as well as the metallicity in the absence of the potential drop or in the low potential field. The dominating magnetization robustly localized along the boundaries with antiferromagnetically correlated edges in the absence of field and an insulating to half-metallic phase transition under low field are observed. These features well reproduce the previous results found experimentally~\cite{Magda-Nature2014} or predicted theoretically~\cite{Son-Nature2006,Son-PRL2006}, manifesting the validity of present method.
\bigskip

\textbf{Model simulations.} We further simulate the magnetic features by the $\pi$-orbital Hubbard model
\begin{equation}
H=-\sum_{ij\sigma}t_{ij}c^{\dag}_{\sigma}c_{j\sigma}+U\sum_{i}n_{i\uparrow }n_{i\downarrow}+\sum_{i\sigma}V_{i} n_{i\sigma}-\mu
\sum_{i}n_{i},
\end{equation}
where $c_{i\sigma}$ is the electron annihilation operator with spin index $\sigma=\pm 1$ at site $i$,
$n_{i\sigma}=c_{i\sigma}^{\dagger}c^{}_{i\sigma}$ the electron number operator. $t_{ij}$ the tight-binding hopping parameters, $U$ the on-site Coulomb repulsion, and $\mu$ the chemical potential. The on-site Coulomb repulsion is set as $U=2t$ unless specified, below the critical value for magnetism in monolayer graphene~\cite{Fujita-JPSJ1996}. In fact, our main conclusions are insensitive to value of $U$. In the absence of external field, this effective model has been shown to be able to capture the main low-energy physics of graphene, and well reproduces the magnetic properties on nanoribbons observed in experiments\cite{Magda-Nature2014} and numerical calculations\cite{Chen-NL2017}. Under low potential drop, the robust edge spin polarization and the insulating to half-metallic phase transition are observed, well reproducing the previous first-principles simulations in the homogeneously low field~\cite{Son-Nature2006}, manifesting the validity of present model. We fix $T=0.01t$ (about $300$ K), for which the room-temperature realizable magnetism promises the future applications. The potential field is modeled by $V_{i}=\frac{V}{2}\frac{e^{-\alpha x_{i}}-e^{\alpha (x_{i}-1)}}{1-e^{-\alpha}}$ with $x_{i}=\frac{3i-3}{3N-2}$ and $\frac{3i-2}{3N-2}$ the normalized positions for $A$ and $B$ sublattice along the finite direction of the ribbon, respectively. We introduce an additional parameter $\alpha$ to simulate the decay of the potential drop in the bulk. In the extend-$\pi$-orbital Hubbard model, the tight-binding parameters are set as $t=2.7$ eV, $t^{\prime}=0.2$ eV, and $t^{\prime\prime}=0.18$ eV according to the first-principles simulations~\cite{Kundu-MPLB2011}. We also use a simplified version--the simple-$\pi$-orbital Hubbard model--by including only the nearest-neighbor hopping parameter $t=2.7$ ev while neglecting all long-range processes.
\bigskip

\textbf{Acknowledgement} We thank WG Yin for helpful discussions and suggestions. This work is supported by the National Natural Science Foundation of China Grant Nos. 12074175. YZ and CDG also acknowledge the Ministry of Science and Technology of China under Grant No. 2016YFA0300401.

\newpage
\setcounter{figure}{0}
\setcounter{equation}{0}
\setcounter{subsection}{0}
\renewcommand \thefigure{S\arabic{figure}}
\renewcommand \theequation{S\arabic{equation}}

\section*{Supplementary information}

\subsection{Magnetism of ZGNR under potential field} 
\subsubsection{Low potential drops} \label{S1A}
\begin{figure}[tbp!]
\includegraphics[width=\columnwidth]{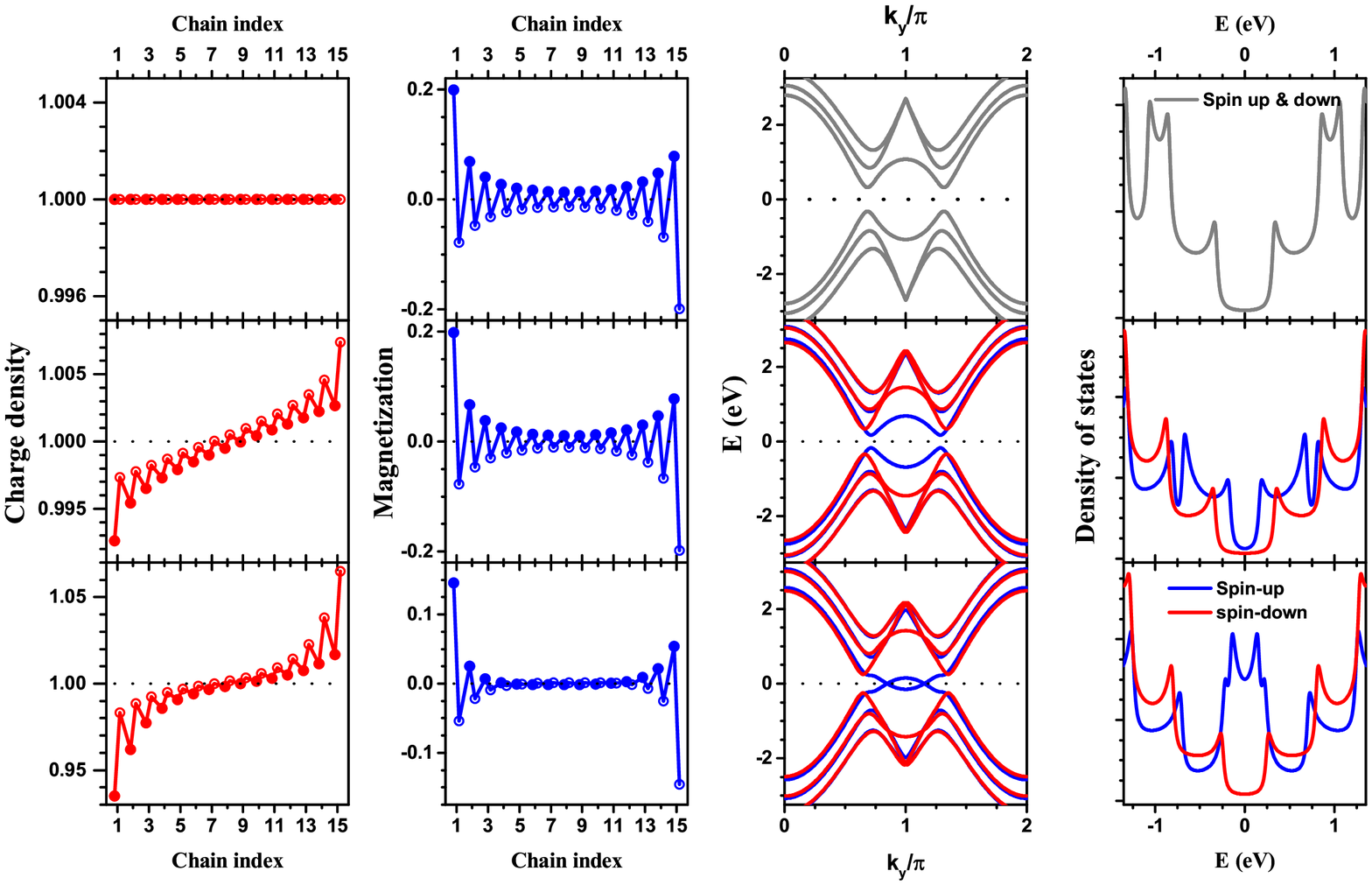} 
\caption{Results in low potential drop of $15$-ZGNR in the framework of simple-$\pi$-orbital model simulations. From left to right panels are the charge density, magnetization, low-energy dispersion (only three lowest conducting and three highest valence bands are included), and the density of states, respectively. From top to bottom, the selected potential drop is $V=0$ eV, $V=0.81$ eV, and $V=1.6$ eV, respectively. The homogeneous potential drop ($\alpha=0$) is assumed. }
\label{Sup-LowField}
\end{figure}

\begin{figure}[tbp!]
\includegraphics[width=0.8\columnwidth]{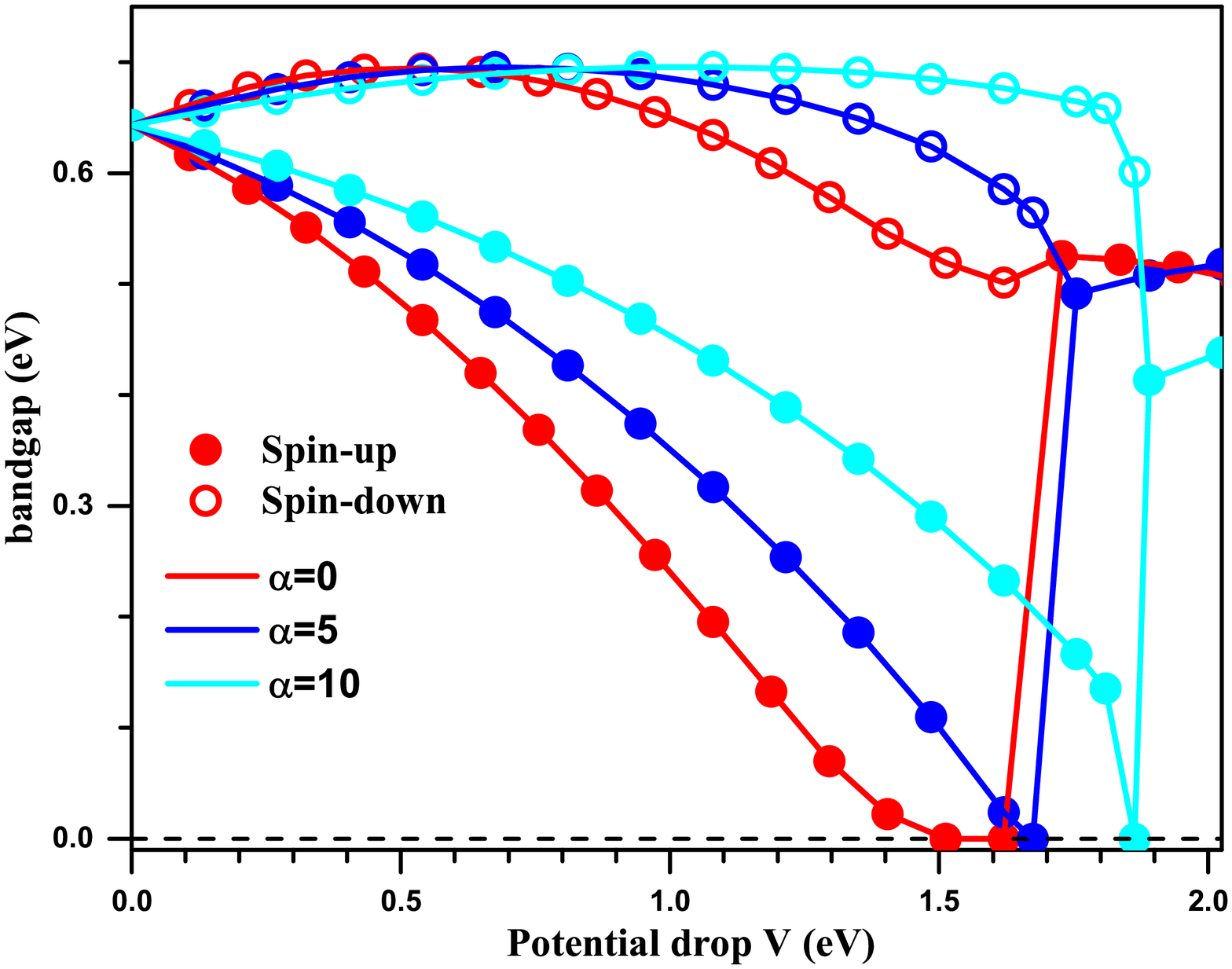} 
\caption{Evolution of the bandgap for spin-up and spin-down channel in low potential drops, simulated by the simple-$\pi$-orbital Hubbard model.}
\label{Sup-Bandgap}
\end{figure}

The typical results in the low potential drop are shown in Fig.~\ref{Sup-LowField}. The spin polarization mainly locates along two edges with the antiferromagnetic correlation between the two edges. In the framework of simple-$\pi$-orbital model, the dispersion of spin-up and spin-down channels is degenerated due to the sublattice symmetry in the absence of the potential drop. Concomitantly, a substantial band gap opens due to the weak antiferromagnetism\cite{Son-PRL2006}. The degeneration is destroyed in the presence of potential drop. However, the antiferromagnetically correlated edge magnetization remains robust. The band gap of one spin channel (here is spin-up channel) decreases gradually, while another (spin-down) channel changes little. In strong enough potential drop, the band gap of spin-up channel closes, yielding a half-metal state. The observed insulator to half-metal phase transition (Fig.~\ref{Sup-Bandgap}) qualitatively agrees with the previous \textit{ab initio} pseudopotential density function method within the local spin density approximation\cite{Son-Nature2006}. Further increasing the potential drop, the ribbon system turns into an insulating state but in the absence of magnetization, i.e., a paramagnetic insulator. These features are insensitive to the detailed potential attenuation ratio $\alpha$ as shown in Fig.~\ref{Sup-Bandgap}.

\subsubsection{Strong potential drops} \label{S1B}
\begin{figure}[tbp!]
\includegraphics[width=\columnwidth]{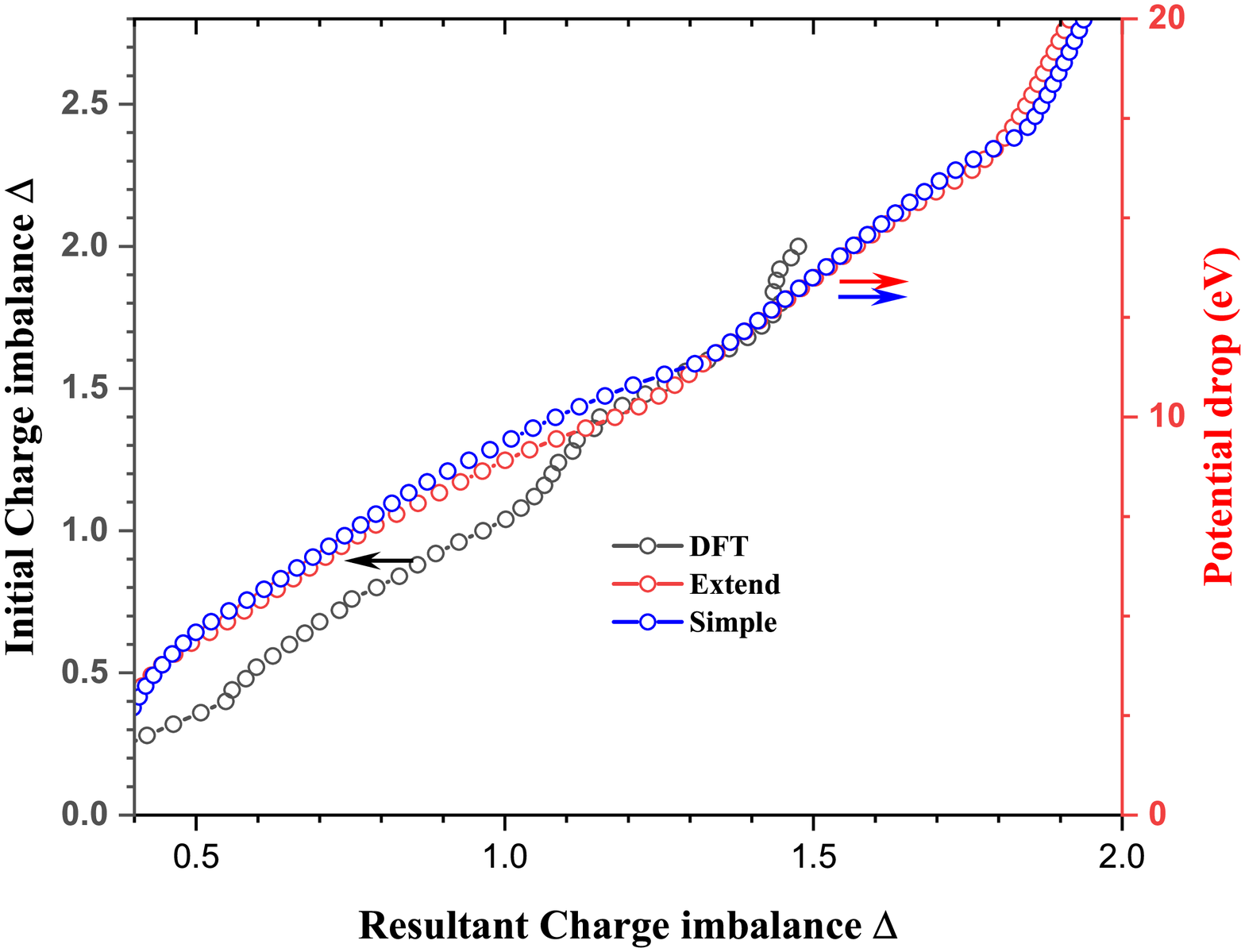} 
\caption{The induced charge imbalance $\Delta$ between the two terminations simulated by ($x$ axis) by the initial virtual crystal approximation in the first-principles calculations (left $y$ axis) and the potential drop in the tight-binding models (right $y$ axis). Here, the spin degree of freedom is neglected. The potential drop and initial charge imbalance are assumed to be linear.}
\label{Imbalance}
\end{figure}

To begin, we compare the influence of the virtual crystal field, in which the linear charge difference with a fixed charge imbalance between the two terminations is initialed, and the potential drop on the charge density distributions. The resultant charge imbalance (between the two terminations) in the first-principles and tight-binding model simulations is show in Fig.~\ref{Imbalance}. The extend- and simple-$\pi$-orbital model simulations show the similar dependence, consistent with the much lower long-range hopping processes in the realistic graphene nanoribbons. On the other hand, the virtual crystal field approximation shows slightly different evolution. In particular, to obtain the resultant charge imbalance $\Delta>1.5$, the initial charge imbalance $\Delta>2$, indicating the $\sigma$ electron of carbon atoms plays role in the virtual crystal field approximation with strong initial charge imbalance. Comparing with the evolution of the resultant charge imbalance $\Delta$ in the first-principles and tight-binding model simulations, each $0.1$ of the initial charge imbalance of the virtual crystal field approximation roughly corresponds to $6\sim 8$ eV of the potential drop.

\begin{figure}[tbp!]
\includegraphics[width=0.8\columnwidth]{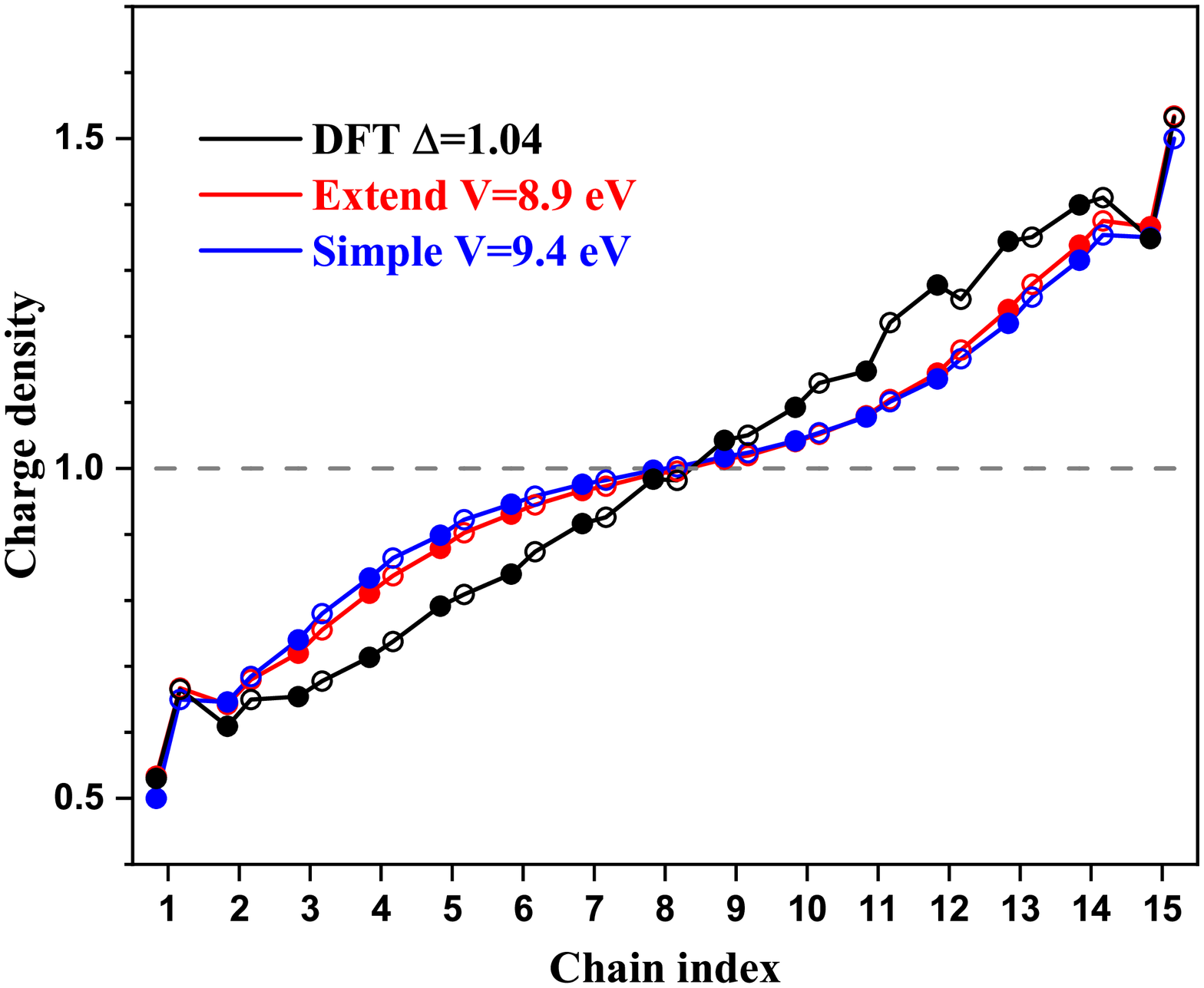} 
\caption{The charge distribution along the finite direction of $15$-ZGNR simulated by the first-principles calculations and the tight-binding models. Here, the resultant charge imbalance is fixed at $\Delta=1$. The potential drop and initial charge imbalance are assumed to be linear.}
\label{DCWD}
\end{figure}

We also show the charge distribution in bulk in the normal state simulated by both the first-principles and tight-binding simulations at the specific resultant charge imbalance $\Delta=1.0$ in Fig.~\ref{DCWD}. The two tight-binding model simulations show almost the same evolution. Similar behavior is also found in the first-principles simulations. It seems that the screening effect is slightly robust in the former cases, in agreement with the phase diagram shown in main text, where the ferromagnetic switches in the latter case is more frequent. 

\begin{figure}[tbp!]
\includegraphics[width=0.8\columnwidth]{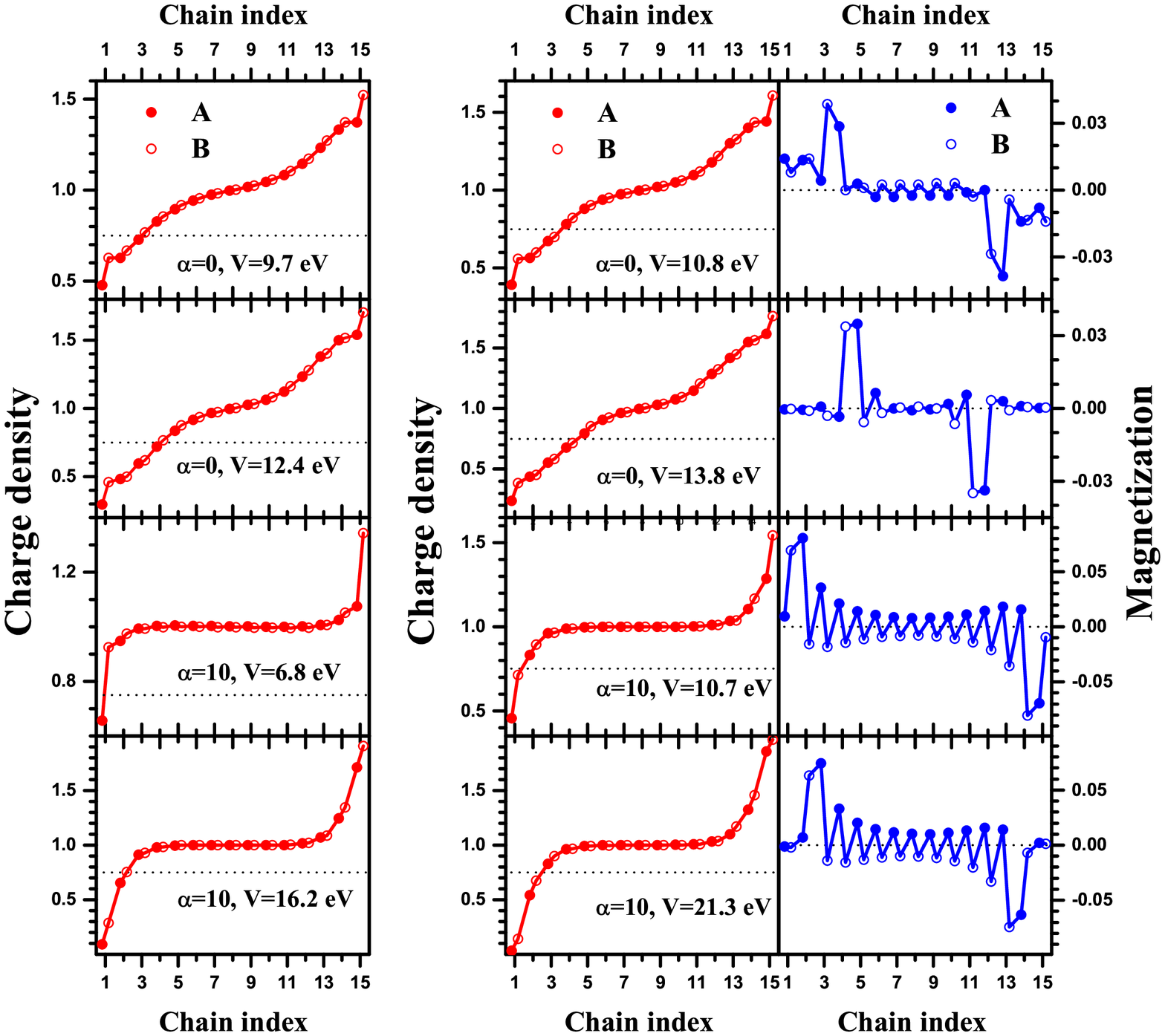} 
\caption{Charge and magnetization distribution in $15$-ZGNR under various potential drops simulated by the simple-$\pi$-orbital Hubbard model. Left panels: the intra-chain state, only charge are shown. From top to bottom, the $3^{\text{rd}}$ intra-chain state ($V=9.7$ eV, $\alpha=0$); the $4^{\text{th}}$ intra-chain state ($V=12.4$ eV, $\alpha=0$); the $1^{\text{st}}$ intra-chain state ($V=6.8$ eV, $\alpha=10$); and the $2^{\text{nd}}$ intra-chain state ($V=16.2$ eV, $\alpha=10$). Black dotted lines denotes the location of Van Hove filling ($\frac{3}{4}$) by linear extrapolation. Right panels: the inter-chain state. From top to bottom, the $3^{\text{rd}}$ inter-chain state ($V=10.8$ eV, $\alpha=0$); the $4^{\text{th}}$ inter-chain state ($V=13.8$ eV, $\alpha=0$); the $1^{\text{st}}$ inter-chain state ($V=10.7$ eV, $\alpha=10$); and the $2^{\text{nd}}$ inter-chain state ($V=21.3$ eV, $\alpha=10$).}
\label{Sup-Orders}
\end{figure}

\begin{figure}[tbp!]
\includegraphics[width=\columnwidth]{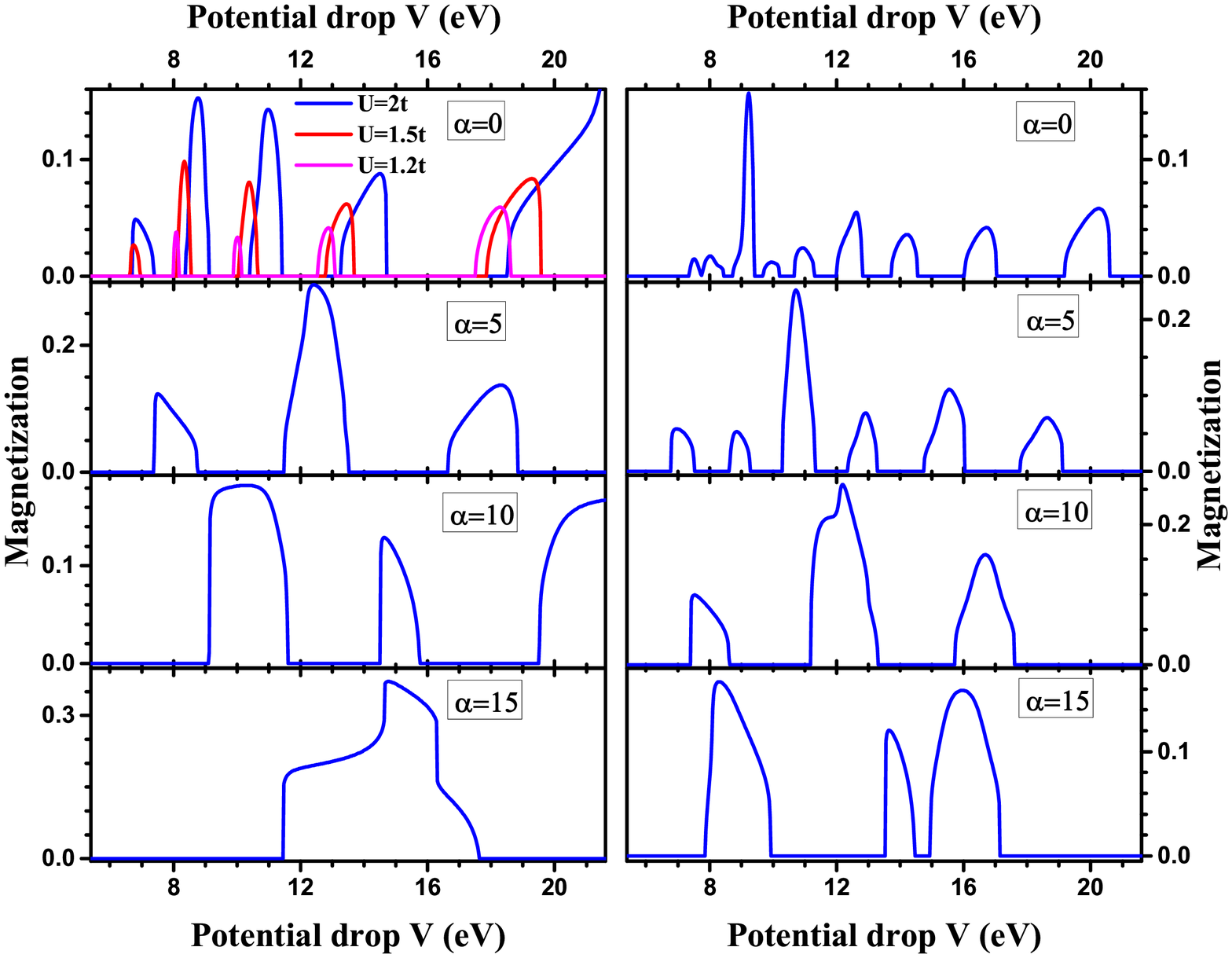} 
\caption{Potential-tuned ferromagnetic switches. Left panels are for the $15$-ZGNR, and right panels are for the $30$-ZGNR. From top to bottom, the decay ratio $\alpha$ is $0$, $5$, $10$, and $15$, respectively. The data with weak $U=1.5t$ ($4.05$ eV) and $U=1.2t$ ($3.24$ eV) in the $15$-ZGNR is also shown in the homogeneous field $\alpha=0$. The magnetization of ribbon is defined as the summation of magnetization on the left half-width (hole-doped side) $M_{sum}=\sum_{i=1}^{N/2} (M_{i}^{A}+M_{i}^{B})$ with $N$ the ribbon width.}
\label{Sup-Switches}
\end{figure}

Fig.~\ref{Sup-Orders} shows some typical distributions of the charge and magnetization in the intra-chain and inter-chain states with different potential drops $V$ and the decay ratio $\alpha$ in $15$-ZGNR in the framework of the simple-$\pi$-orbital Hubbard model. The magnetic phase emerges only in the $i^{\text{th}}$ inter-chain state with dominating ferromagnetic domain at the inter-chain atoms $B_{i}$ and $A_{i+1}$, irrespective of the details. The robust switches of the magnetism are shown in Fig.~\ref{Sup-Switches}. The magnetic switches upon the potential drops are insensitive to the decay ratio $\alpha$, width of ribbon $N$, and even the on-site Coulomb repulsion $U$, indicating their intrinsic nature.

\begin{figure}[tbp!]
\includegraphics[width=\columnwidth]{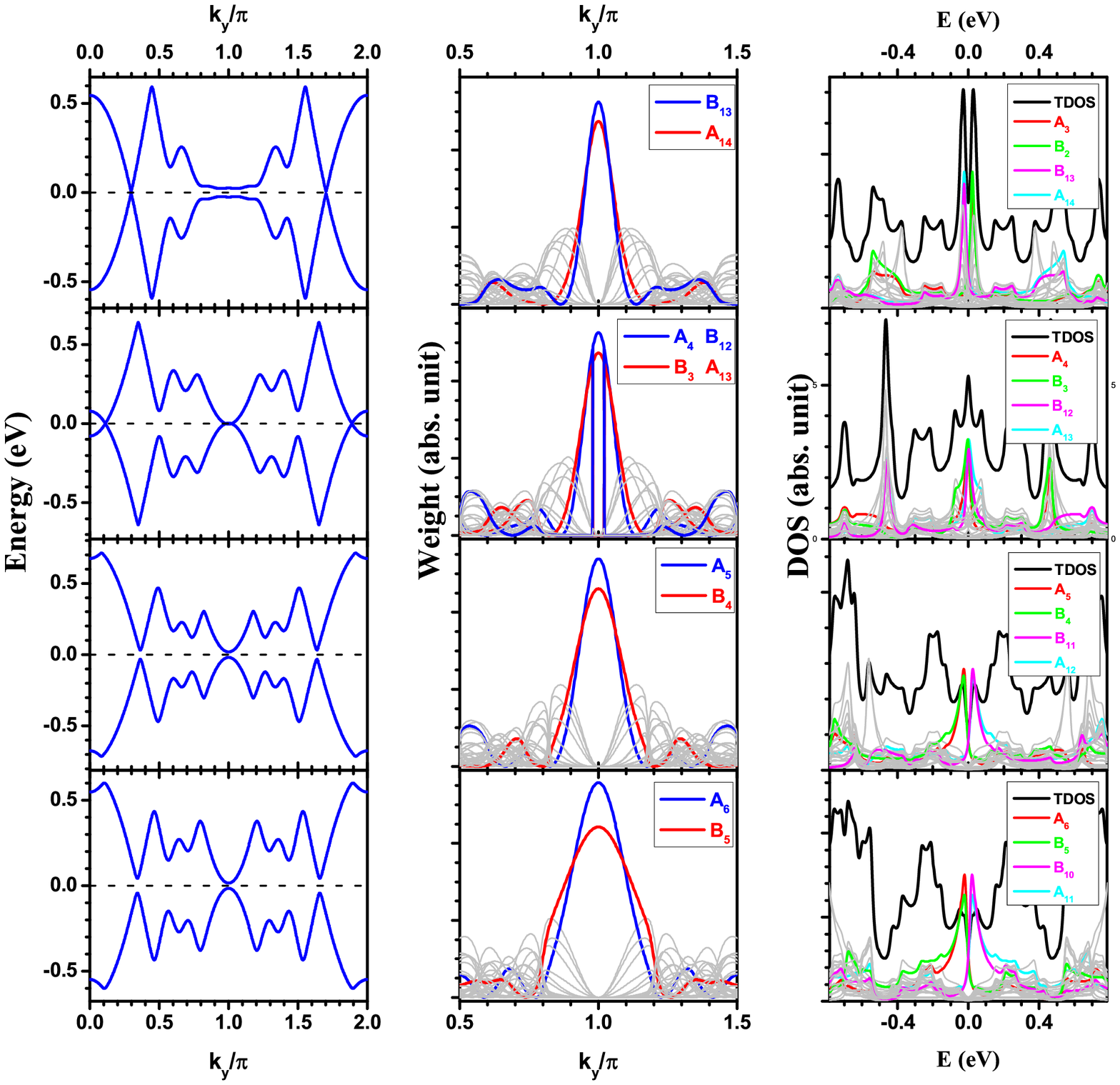} 
\caption{Low energy band structure and density of states in the intra-chain states simulated by the simple-$\pi$-orbital Hubbard model. Left panels are the low energy bands, for simplicity, only the lowest conducting and the highest valence band are depicted. Middle panels are the spectral wights of the respective sites for the highest valence band shown in left panels. The main contribution from the inter-chain sites for the near $\pi$-region bands are highlighted. Right panels are the density of states for the respective intra-chain state shown in the left and middle panels. The local density of states for respective sites are also shown. The corresponding local density of states of the sites specified in the middle panels are also highlighted, together with their particle-hole counterpart. Note, the total density of states (TDOS in right panels) have been divided by a factor $4$. From the top to bottom, $V=9.5$ eV (the $3^{\text{rd}}$ intra-chain state), $V=12.7$ eV (the $4^{\text{th}}$ intra-chain state), $V=14.9$ eV (the $5^{\text{th}}$ intra-chain state, and $V=17.6$ eV (the $5^{\text{th}}$ intra-chain state), respectively. The selected system is the $15$-ZGNR with the homogeneous potential drop ($\alpha=0$), $U=5.4$ eV.}
\label{Sup-BandDOS-intra}
\end{figure}

\begin{figure}[tbp!]
\includegraphics[width=\columnwidth]{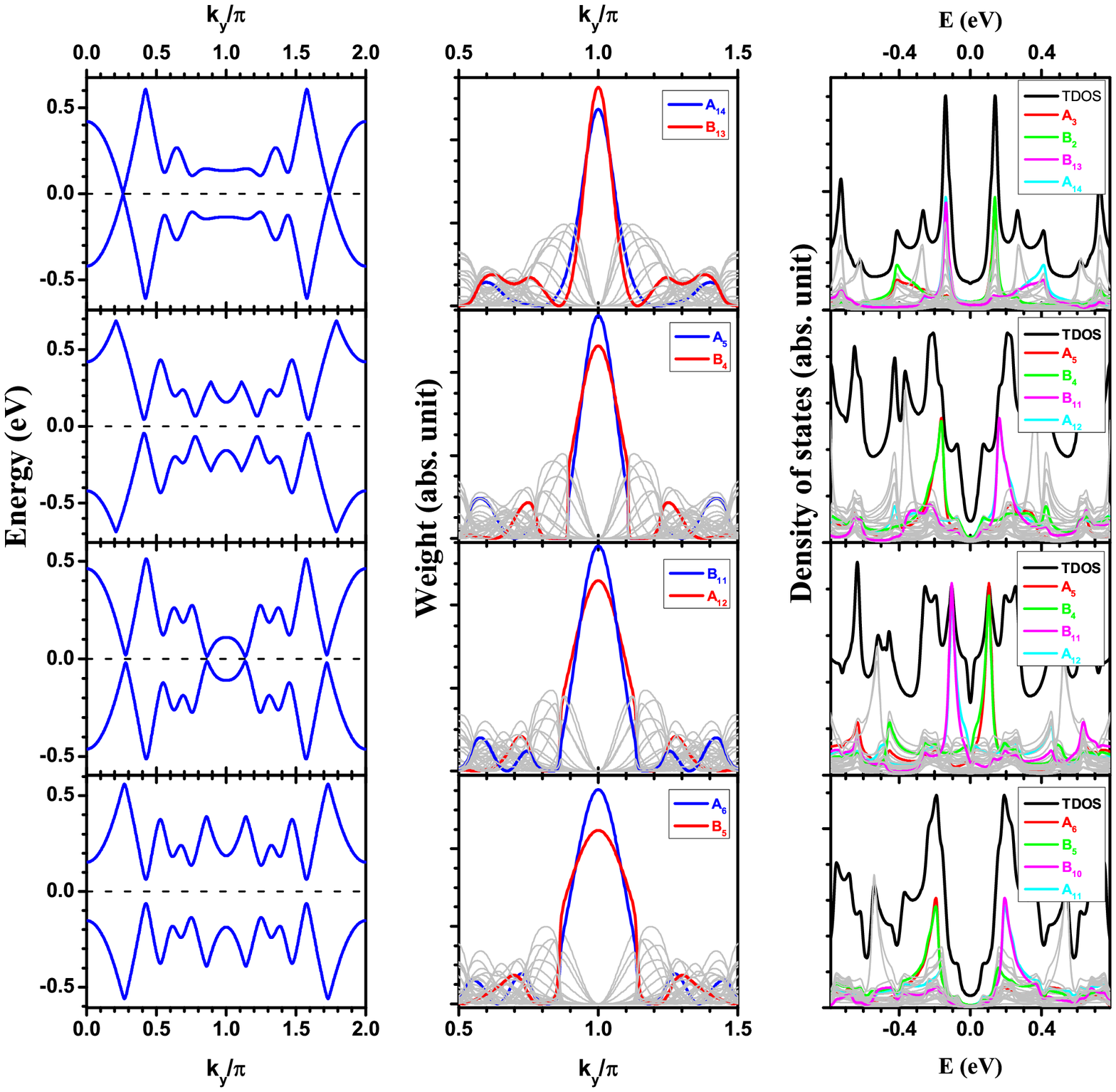} 
\caption{Low energy band structure and density of states in the inter-chain states simulated by the simple-$\pi$-orbital Hubbard model. The notations are same as Fig.~\ref{Sup-BandDOS-intra}. From the top to bottom, $V=8.9$ eV (the $2^{\text{nd}}$ intra-chain state), $V=10.8$ eV (the $3^{\text{rd}}$ intra-chain state), $V=13.5$ eV (the $4^{\text{th}}$ intra-chain state, and $V=18.9$ eV (the $5^{\text{th}}$ intra-chain state), respectively. The selected system is the $15$-ZGNR with the homogeneous potential drop ($\alpha=0$), $U=5.4$ eV.}
\label{Sup-BandDOS-inter}
\end{figure}

\begin{figure}[tbp!]
\includegraphics[width=0.8\columnwidth]{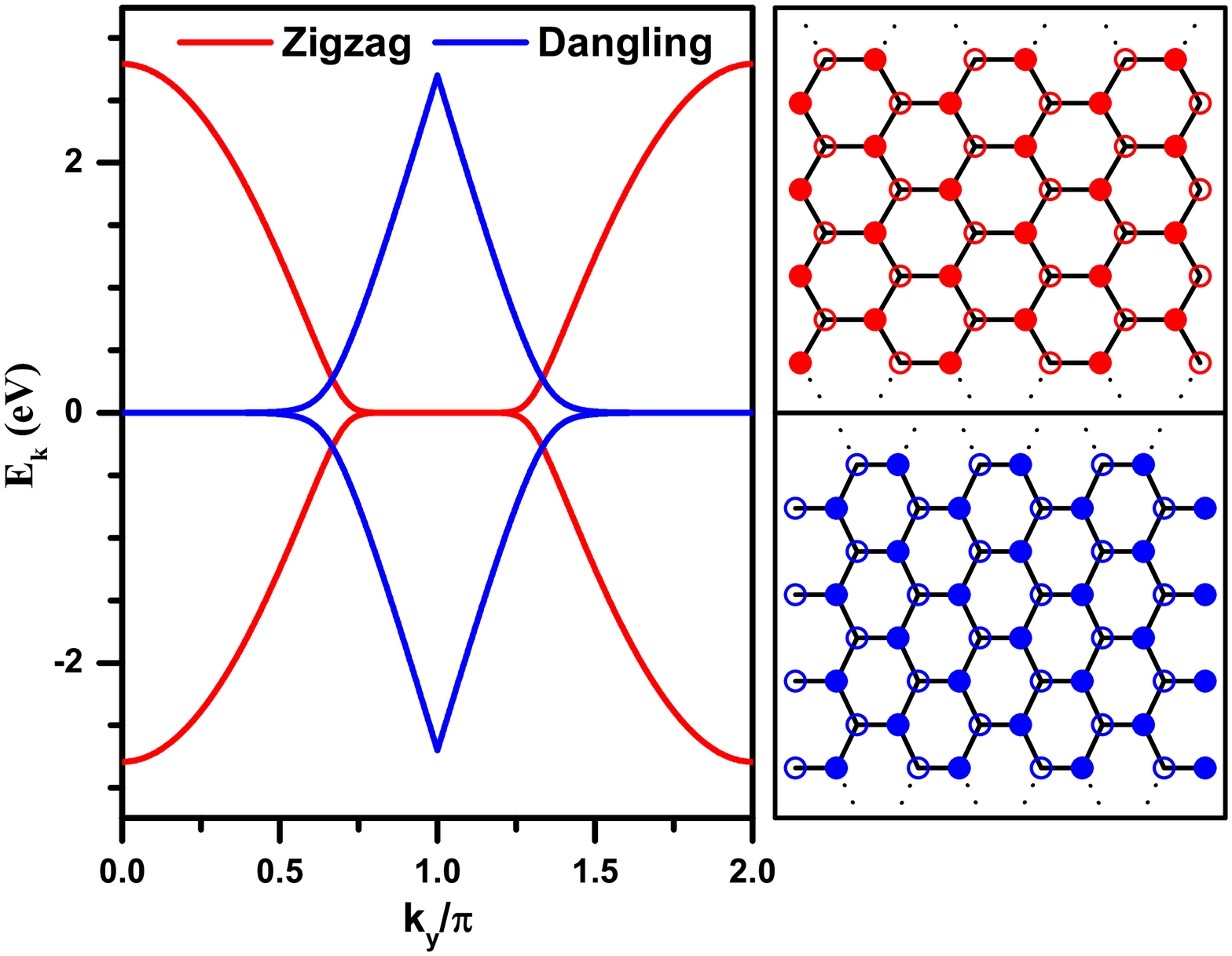} 
\caption{Left panel: Lowest conducting and highest valence band in the absence of the potential drop in zigzag- and dangling-edge graphene nanoribbons. Right panels are two schematic graphene ribbons, upper panel is for the zigzag-, and low panel is for the dangling-edge ribbon, respectively.}
\label{Sup-Shape}
\end{figure}

We show the low-energy band structures of the intra- (Fig.~\ref{Sup-BandDOS-intra}) and inter-chain states (Fig.~\ref{Sup-BandDOS-inter}) in normal state. The narrow bands near the Fermi level emerge in both states, substantially enhancing the strong correlations. The gap between the lowest conducting and the highest valence band is almost closed near $\pi$ region in the inter-chain states, in sharp contrast with the robust band gap in the intra-chain states. Such difference provides the large DOS at Fermi energy in the inter-chain states, while negligible DOS at Fermi energy in the intra-chain states. Therefore, zigzag edge induced strong correlations, together with a large DOS near the Fermi energy in the inter-chain states. They are two of three essential ingredients for the ordered state. As we argued in the main text, the virtual edges where Van Hove fillings locates are dangling and zigzag edges in the intra- and inter-chain states. The gap opening and closing near the $\pi$ region in the intra- and inter-chain states is well consistent with the respective graphene nanoribbons as schematically illustrated in Fig.~\ref{Sup-Shape}.

\subsection{Half-metallicity} \label{S2}
\begin{figure}[tbp!]
\vspace{-0in} \hspace{-0.0in} \center
\includegraphics[width=\columnwidth]{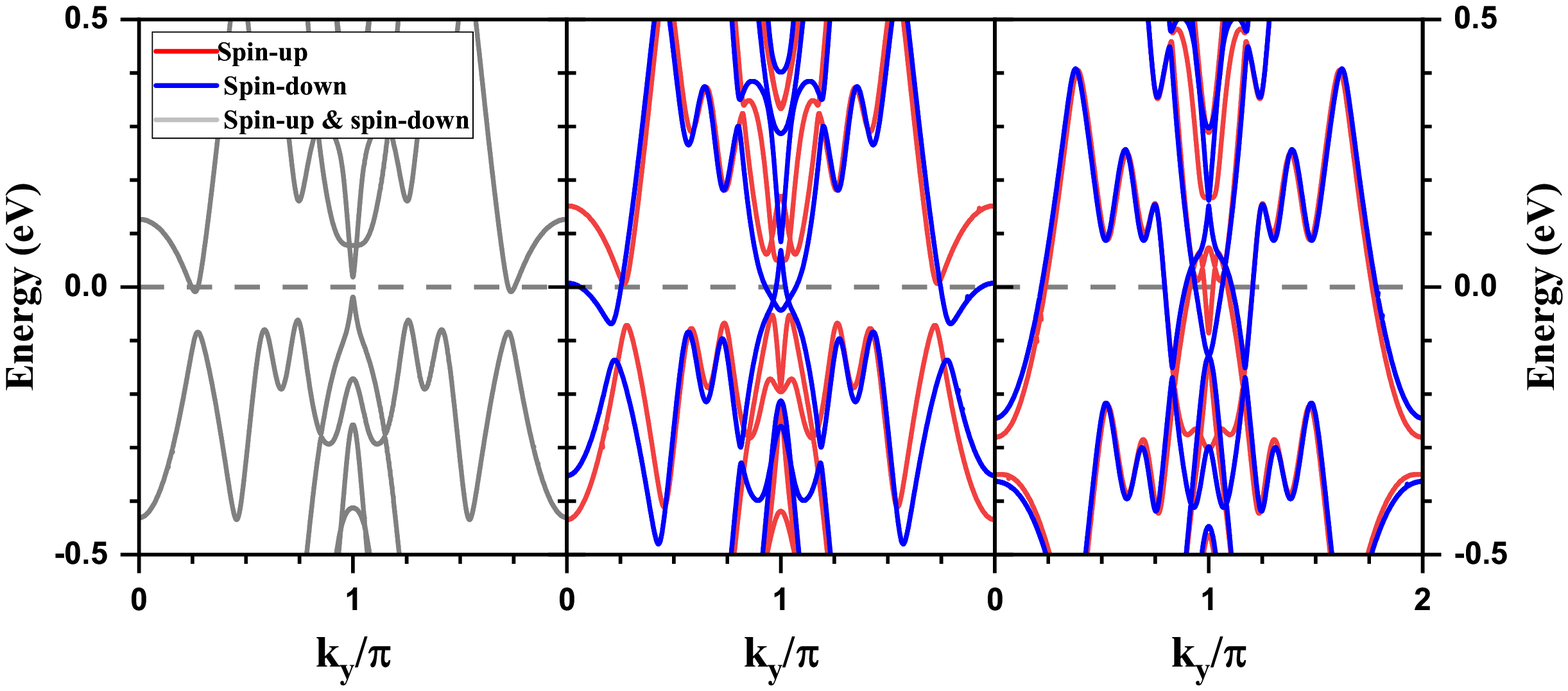} 
\caption{Spin resolved band structure evolution simulated by the first-principles calculations. Left panel is for the $2^{\text{nd}}$ intra-chain state ($\Delta=0.72$), in which energy band in the spin-up and spin-down channels are degenerated. Middle ($\Delta=0.8$) and right ($\Delta=1.0$) panels are for the $3^{\text{rd}}$ inter-chain state.}
\label{Sup-HalfDFT}
\end{figure}

We show the band structure of for the two spin channels simulated by the first-principles calculations. We note that the band gaps may close at the Dirac points inherited from the bulk of graphene. In this sense, the half-metal behavior discussed here is more precisely called as the half-metal-like behavior. Fig.~\ref{Sup-HalfDFT} shows the spin-resolved band structure, corresponding to the spin-resolved density of states shown in Fig.3 (a) in main text, in details. In the $2^{\text{nd}}$ intra-chain state (left panel), the ribbon system is an insulating paramagnetic state. Entering the $3^{\text{rd}}$ ($V=0.2$), the half-metal-like behavior with the band gap opening in the spin-down channel while closing in the spin-up channel is observed. Further increasing the charge imbalance, the band gaps close in both spin channels, resulting in a metallic behavior.

\begin{figure}[tbp!]
\vspace{-0in} \hspace{-0.0in} \center
\includegraphics[width=\columnwidth]{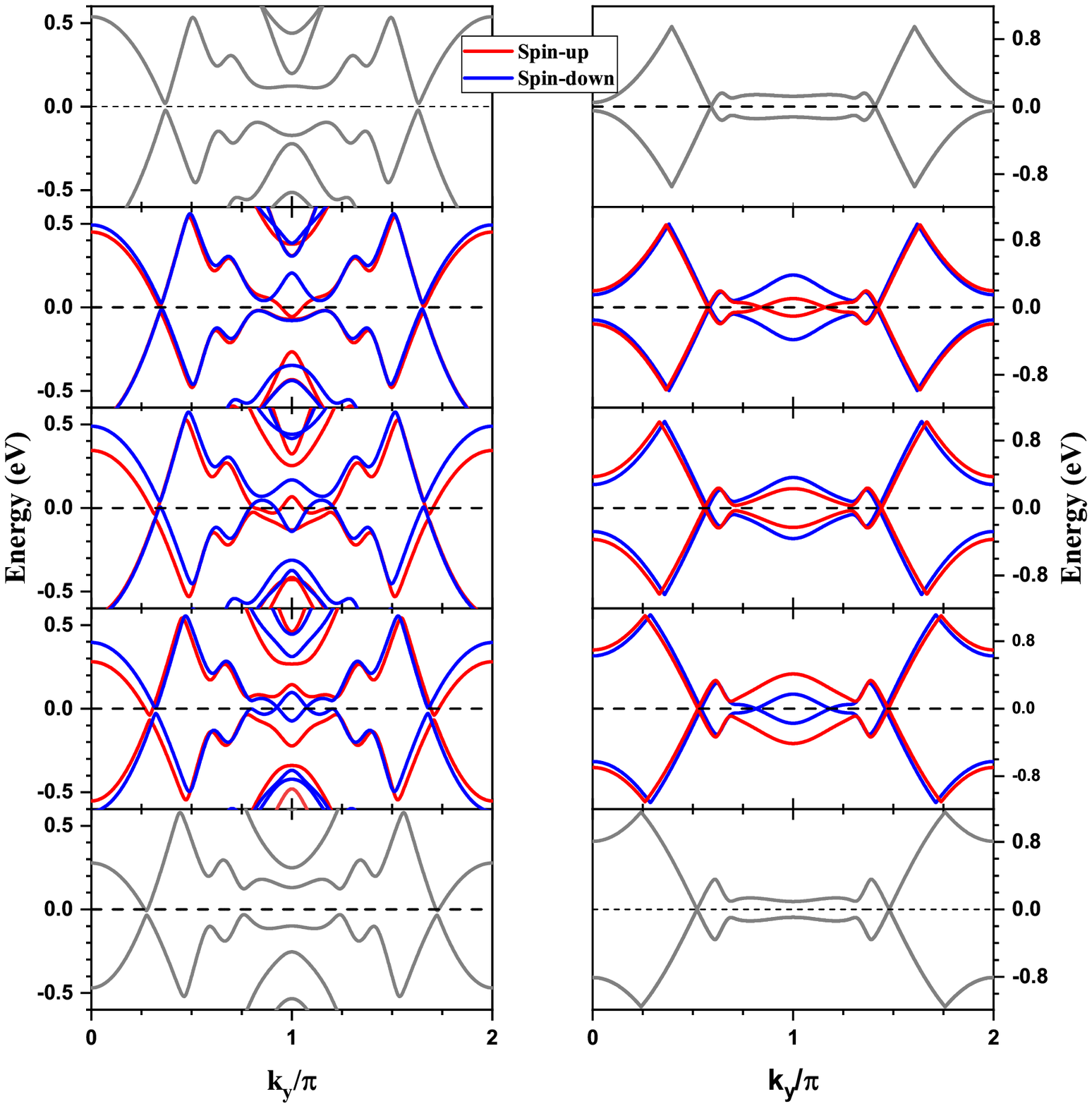} 
\caption{Spin resolved band structure evolution in the $2^{\text{nd}}$ inter-chain state. Left panels are for the results simulated by the extend-$\pi$-orbital Hubbard model in the $15$-ZGNR. The homogeneous potential drops are assumed ($\alpha=0$). From top to bottom, $V=7.0$, $7.3$, $7.6$, $8.0$ and $8.4$ eV, respectively, corresponding to the spin resolved density of state shown in Fig. 3 (b) in main text. Right panels are for the results simulated by the simple-$\pi$-orbital Hubbard model in the $30$-ZGNR with the strong decayed potential drop ($\alpha=15$). From top to bottom, $V=14.6$, $15.1$, $15.7$, $16.7$, and $17.3$ eV, respectively, corresponding to the bandgap of the spin-up and spin-down channels depicted in Fig. 3 (c) in main text. The panels with gray lines are in the intra-chain paramagnetic state, in which the band of spin-up and spin-down channels are degenerate.}
\label{Sup-Halfmetal}
\end{figure}

We further show the band structure for the two spin channels simulated by both the extend- and simple-$\pi$-orbital Hubbard model. In the left panels of Fig.~\ref{Sup-Halfmetal}, we show the evolution of the spin-resolved band gap from the $2^{\text{nd}}$ to the $3^{\text{rd}}$ intra-chain state simulated by the extend-$\pi$-orbital Hubbard model. The band gap opens near the $k_{y}=\pi$ region in the intra-chain state. As a result, the relatively low density of states near the Fermi energy leads to a paramagnetic state as mentioned in the main text. The bands of the spin-up and spin-down channels are degenerate. Entering the $2^{\text{nd}}$ inter-chain state, the band gap of the spin-down channel remain opens, whereas that of the spin-up channel closes, resulting in a half-metal (spin-up channel) behavior. With the increase of the potential drop, the band gap closes in the spin-down channel, while opens in the spin-up channel, again leading to a half-metal feature but with reversed spin channel. We, therefore, find an interesting half-metallicity transition between the reversal spin channels by tuning the potential drop. Such transition provides new route to manipulate the spin transport properties. Similar transition of the half-metallicity is also observed in the simple-$\pi$-orbital Hubbard model simulations in wider ribbons ($N=30$) as shown in the right panels of Fig.~\ref{Sup-Halfmetal}, manifesting the robustness of the discovered half-metallicity transition in the present potential-tuned ZGNRs.

\subsection{Spin dynamics in graphene nanoribbon} \label{S3}
We consider the non-interacting system ($U=0$) in the potential field. Since the spin-up and spin-down channels are degenerate in the absence interaction, we neglect the spin index. The Hamiltonian in half-momentum space (open condition along the $x$ direction and periodic along the $y$ direction) is
\begin{eqnarray}
H &=&-\sum_{i,k}\left[ \left(
\begin{array}{c}
\left( t+2t^{\prime \prime }\cos
k_{x}\right) a_{i,k}^{\dag }b_{i-1,k} \nonumber \\
+t\gamma _{k}a_{i,k}^{\dag }b_{i,k}+t^{\prime \prime }e^{-ik}a_{i,k}^{\dag }b_{i+1,k}%
\end{array}%
\right) +h.c.\right]  \\
&-&\sum_{i,k}\left[ 2t^{\prime}\cos k_{x}a_{i,k}^{\dag }a_{i,k}+t^{\prime}\left( \gamma
_{k}a_{i,k}^{\dag }a_{i+1,k}+h.c.\right) \right] \nonumber \\
&-&\sum_{i,k}\left[ 2t^{\prime}\cos k_{x}b_{i,k}^{\dag }b_{i,k}+t^{\prime}\left( \gamma
_{k}b_{i,k}^{\dag }b_{i+1,k}+h.c.\right) \right].
\end{eqnarray}
Here, $a_{i,k}/b_{i,k}$ is the electron annihilation operator in the $A/B$ sublattice. $i=1,N$, and we use the notation $A/B_{i\pm 1\notin[1,N]}=0$.
$\gamma _{k}=\left( 1+e^{-ik}\right) $ with the distance between the
nearest-neighbor atoms in the same sublattice being the unit. The tight-binding order parameters are $t=2.7$ eV, $t^{\prime}=0.2$ eV and $t^{\prime\prime}=0.18$ eV for the extend-$\pi$-orbital model, and $t=2.7$ eV, $t^{\prime}=t^{\prime\prime}=0$ for the simple-$\pi$-orbital model. The Hamiltonian can be diagonalized by introducing the rotational
transformation%
\begin{equation}
a_{i,k}=\sum_{j=1}^{2N}\mathcal{T}_{2i-1,j}\alpha _{j,k},\text{ and }%
b_{i,k}=\sum_{j=1}^{2N}\mathcal{T}_{2i,j}\alpha _{j,k},
\end{equation}
where $\mathcal{T}_{i,j}$ is the element of the matrix $\mathcal{T}%
_{2N\times 2N}$, $\alpha _{j,k}$ is the quasiparticle operator. Taking
the basis $\Psi =\left( a_{1,k },b_{1,k },\cdots ,a_{N,k
},b_{N,k }\right) $, the Hamiltonian in quasiparticle
representation is expressed as%
\begin{equation}
H=\sum_{k}\sum_{\nu =1}^{2N}E_{k}^{\nu }\alpha _{\nu, k}^{\dag }\alpha _{\nu,
k}
\end{equation}
with $E_{k}^{\nu }$ the energy of the quasiparticle. The Green's function is
then defined as
\begin{eqnarray}
\mathcal{G}_{i,j}^{a,a}\left( k\right)  &=&\left\langle a_{i,k}a_{j,k}^{\dag
}\right\rangle =\sum_{\nu =1}^{2N}\frac{\mathcal{T}_{2i-1,\nu }\mathcal{T}%
_{2j-1,\nu }^{\ast }}{i\omega _{n}-E_{k}^{\nu }}, \nonumber \\
\mathcal{G}_{i,j}^{b,b}\left( k\right)  &=&\left\langle b_{i,k}b_{j,k}^{\dag
}\right\rangle =\sum_{\nu =1}^{2N}\frac{\mathcal{T}%
_{2i,\nu }\mathcal{T}_{2j,\nu }^{\ast }}{i\omega _{n}-E_{k}^{\nu }}, \nonumber \\
\mathcal{G}_{i,j}^{a,b}\left( k\right)  &=&\left\langle a_{i,k}b_{j,k}^{\dag
}\right\rangle =\sum_{\nu =1}^{2N}\frac{\mathcal{T}%
_{2i-1,\nu }\mathcal{T}_{2j,\nu }^{\ast }}{i\omega _{n}-E_{k}^{\nu }}, \nonumber \\
\mathcal{G}_{i,j}^{b,a}\left( k\right)  &=&\left\langle b_{i,k}a_{j,k}^{\dag
}\right\rangle =\sum_{\nu =1}^{2N}\frac{\mathcal{T}%
_{2i,\nu }\mathcal{T}_{2j-1,\nu }^{\ast }}{i\omega _{n}-E_{k}^{\nu }}.
\end{eqnarray}

We calculate four types of transverse spin susceptibility defined as%
\begin{eqnarray}
\chi _{i,i}^{a,a}\left( q,i\omega _{m}\right)  &=&-\frac{1}{N_{y}}\sum_{k}\left(
k\right) \mathcal{G}^{aa}_{ii}\left( k\right) \mathcal{G}_{ii}^{aa}\left(
k+q\right)  \nonumber\\
\chi _{i,i}^{b,b}\left( q,i\omega _{m}\right)  &=&-\frac{1}{N_{y}}\sum_{k}\mathcal{G}%
_{ii}^{bb}\left( k\right) \mathcal{G}_{ii}^{bb}\left( k+q\right)  \nonumber\\
\chi _{i,i}^{a,b}\left( q,i\omega _{m}\right)  &=&-\frac{1}{N_{y}}\sum_{k}\mathcal{G}%
_{ii}^{ab}\left( k\right) \mathcal{G}_{ii}^{ba}\left( k+q\right)  \nonumber\\
\chi _{i,i+1}^{b,a}\left( q,i\omega _{m}\right)  &=&-\frac{1}{N_{y}}\sum_{k}\mathcal{G}%
_{i,i+1}^{ba}\left( k\right) \mathcal{G}_{i+1,i}^{ab}\left( k+q\right)
\end{eqnarray}

Here, $\omega _{m}$ is the matsubara frequency, $N_{y}$ is the number of $k_{y}$ points along the periodic direction. The momentum dependent spin
correlation is obtained at zero frequency $\omega _{m}=0$. Hence, $\chi_{A_{i}A_{i}}(q,0)$ and $\chi_{B_{i}B_{i}}(q,0)$ represent the self-correlation between the same sublattice, $\chi_{A_{i}B_{i}}(q,0)$ is the intra-chain correlation, and $\chi_{B_{i+1}A_{i}}(q,0)$ is the inter-chain correlation.

\begin{figure}[tbp!]
\vspace{-0in} \hspace{-0.0in} \center
\includegraphics[width=\columnwidth]{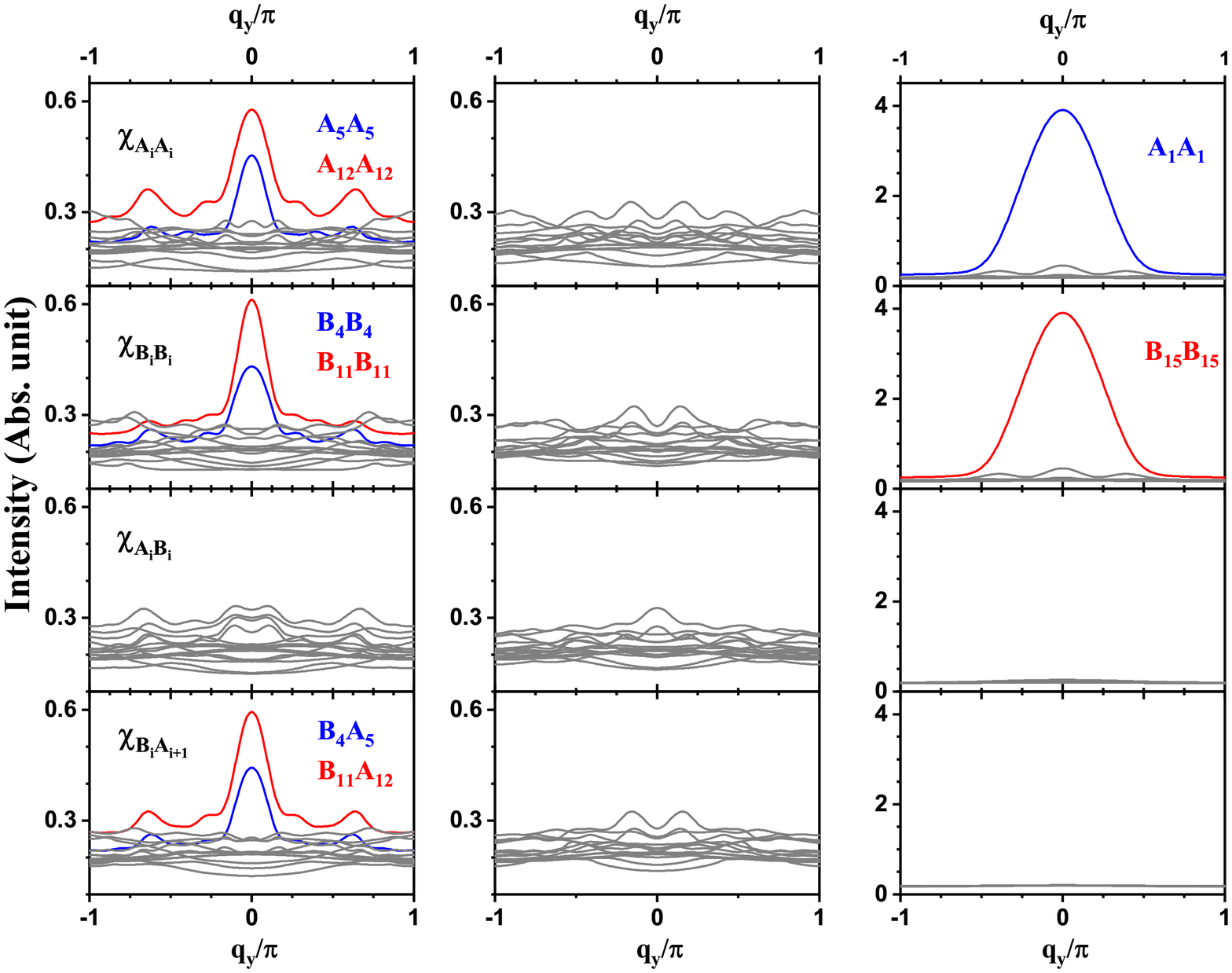} 
\caption{Static spin susceptibility in various situations on the $15$-ZGNRs simulated by the extend-$\pi$-orbital model. The left panels, and middle panels are for the $4^{\text{th}}$ inter-chain state ($V=12.5$ eV), and the $4^{\text{th}}$ intra-chain state ($V=11.5$ eV), respectively. The homogeneous potential drops are assumed ($\alpha=0$).  The right panels are for the case of no potential drop applying ($V=0$).}
\label{Sup-sus}
\end{figure}

To append the spin correlations (Fig.~(1) (c) in main text) discussed in main text, we show the more details of the static spin susceptibility. The results in the $4^{\text{th}}$ inter-chain state is shown in the left panels of Fig.~\ref{Sup-sus}. Prominent peaks at $q_{y}=0$ are found in $\chi^{a,a}_{5,5}$ and $\chi^{b,b}_{4,4}$ components (and their counterparts), and the inter-chain component $\chi^{b,a}_{4,5}$, indicating the strong tendency towards the ferromagnetic ground state in those channels. In comparison, weak or negligible peaks are found in other components. The inter-chain ($4^{\text{th}}$) ferromagnetic domain are, therefore, expected. On the other hand, no evident peaks are observed in all components in the $4^{\text{th}}$ intra-chain state (middle panels in Fig~\ref{Sup-sus}), corresponding to the paramagnetic ground state in the intra-chain state. We also show the results in the absence of potential drop in the right panels of Fig~\ref{Sup-sus}, the extensive peaks at $q_{y}=0$ is found in $\chi^{a,a}_{1,1}$ and $\chi^{b,b}_{15,15}$, while other channels are negligible. This consists with the strong spin polarization along the edges in the absence of potential drop as revealed by previous experimental measurements~\cite{Magda-Nature2014} and theoretical calculations~\cite{Jiang-JCP2007,Ma-PRB2016,Chen-NL2017}.\bigskip

\subsection{Density of state} \label{S4}
According to Supplementary~\ref{S3}, the retard Green's function at given site is
\begin{equation}
\mathcal{G}_{ii\sigma }^{aa}\left( k,i\omega _{n}\right) =\left\langle
a_{i,k,\sigma }a_{i,k,\sigma }^{\dag }\right\rangle =\sum_{\nu =1}^{2N}\frac{%
\left\vert \mathcal{T}_{2i-1,\nu }^{\sigma }\right\vert ^{2}}{i\omega
_{n}-E_{k\sigma }^{\nu }},
\end{equation}
and%
\begin{equation}
\mathcal{G}_{ii\sigma }^{bb}\left( k,i\omega _{n}\right) =\left\langle
b_{i,k,\sigma }b_{i,k,\sigma }^{\dag }\right\rangle =\sum_{\nu =1}^{2N}\frac{%
\left\vert \mathcal{T}_{2i,\nu }^{\sigma }\right\vert ^{2}}{i\omega
_{n}-E_{k\sigma }^{\nu }},
\end{equation}
where $\mathcal{T}_{i,\nu }^{\sigma }$ ($i,v=1,2,\cdots ,2N$) is the element
of the transformation matrix for spin-up and spin-down channel, and $%
E_{k\sigma }^{\nu }$ is the energy of quasiparticle $\alpha _{\nu k\sigma }$%
. The local density of states of the sublattice $A$ and $B$ on chain $i$ is
defined as
\begin{equation}
\mathcal{D}_{A_{i}}\left( \omega \right) =-\frac{1}{\pi }\sum_{k\sigma
}\sum_{\nu =1}^{2N}\Im\left( \frac{\left\vert \mathcal{T}_{2i-1,\nu
}^{\sigma }\right\vert ^{2}}{\omega -E_{k\sigma }^{\nu }+i\Gamma }\right) ,
\end{equation}%
and
\begin{equation}
\mathcal{D}_{B_{i}}\left( \omega \right) =-\frac{1}{\pi }\sum_{k\sigma
}\sum_{\nu =1}^{2N}\Im\left( \frac{\left\vert \mathcal{T}_{2i,\nu
}^{\sigma }\right\vert ^{2}}{\omega -E_{k\sigma }^{\nu }+i\Gamma }\right)
\end{equation}%
with $\Gamma $ a broaden factor. The total density of states is
\begin{equation}
\mathcal{D}\left( \omega \right) \mathcal{=}\frac{1}{2N}\sum_{i=1}^{N}\left(
\mathcal{D}_{A_{i}}\left( \omega \right) +\mathcal{D}_{B_{i}}\left( \omega
\right) \right) .
\end{equation}

\bibliography{ref}

\end{document}